\documentclass[abstract=on,notitlepage,superscriptaddress,twocolumn,prd,prl,nofootinbib,aps,showpac,groupedaddress]{revtex4-2}
\usepackage{srcltx}
\usepackage{epsf}
\usepackage{amsfonts}
\usepackage{amsmath}
\usepackage{mathrsfs}
\usepackage{epsfig}
\usepackage{enumerate}
\usepackage{lipsum}
\usepackage{graphicx}
\usepackage{epsf}
\usepackage{subfig}
\usepackage{color}
\usepackage{caption}
\usepackage{subfig}
\usepackage{epstopdf}
\usepackage{float}
\usepackage{dcolumn}
\usepackage{hyperref}
\usepackage{amsthm}

\usepackage[T1]{fontenc}
\DeclareSymbolFont{TOneChars}{T1}{\familydefault}{m}{it}
\SetSymbolFont{TOneChars}{bold}{T1}{\familydefault}{bx}{it} 
\DeclareMathSymbol{\mathdh}{\mathord}{TOneChars}{"F0}

\newcommand{\be}{\begin{equation}}
\newcommand{\ee}{\end{equation}} 
\newcommand{\eei}{\end{equation}\indent\indent}
\newcommand{\bc}{\begin{center}}
\newcommand{\ec}{\end{center}}
\newcommand{\ber}{\begin{eqnarray*}}
\newcommand{\ear}{\end{eqnarray*}}
\newcommand{\ba}{\begin{array}}
\newcommand{\ea}{\end{array}}

\newcommand{\bea}{\begin{eqnarray}}
\newcommand{\eea}{\end{eqnarray}}

\newcommand{\ei}{\end{itemize}}

\begin{document}


\title{Astrophysical and cosmological scenarios for gravitational wave heating}



\author{Nigel T. Bishop${}^{1,3}$}\email[]{n.bishop@ru.ac.za}
\author{Vishnu Kakkat${}^2$}\email[]{kakkav@unisa.ac.za}
\author{Amos S. Kubeka${}^{2,3}$}\email[]{kubekas@unisa.ac.za}
\author{Monos Naidoo${}^1$}\email[]{m.naidoo@ru.ac.za}
\author{Petrus~J. van der Walt${}^1$}\email[]{p.vanderwalt@ru.ac.za}
\affiliation{${}^1$Department of Mathematics, Rhodes University, Makhanda 6140, South Africa}
\affiliation{${}^2$Department of Mathematical Sciences, Unisa, P.O. Box 392, Pretoria 0003, South Africa}
\affiliation{${}^3$National Institute for Theoretical and Computational Sciences (NITheCS) South Africa}

\begin{abstract}
Gravitational waves (GWs) passing through a viscous shell of matter are expected to be damped resulting in an increase in the temperature of the fluid as energy is transferred to it from the GWs. In previous work we constructed a model for this process, obtaining an expression for the temperature distribution inside the shell, and it was shown that the temperature increase can be astrophysically significant. In this paper we extend the analysis to GW heating and damping following a binary neutron star merger, GW heating during a core-collapse supernova, and primordial gravitational waves.
\end{abstract}
\maketitle

\section{Introduction}

 
The detection of gravitational waves (GWs) from distant events is possible because GWs are usually unaffected by matter, allowing them to traverse cosmological distances without significant attenuation.
This does not mean that GWs do not interact with matter and some of the various ways they do are described in~\cite{hawking1966perturbations,esposito1971absorption,marklund2000radio,brodin2001photon,cuesta2002gravitational}.  Travelling through a perfect fluid, GWs do not experience any absorption or dissipation, however passing through a viscous fluid, GWs transfer energy to the fluid~\cite{hawking1966perturbations} according to 
\begin{equation}
\frac{d\dot{E}_{GW}}{dr}=\frac{16\pi G}{c^3}\eta \dot{E}_{GW}
\label{eqn:eta}
\end{equation}
where $G$ is the gravitational constant, $c$ is the speed of light, $\dot{E}_{GW}$ is the GW power output, $\eta$ is the dynamic viscosity and $r$ is distance. 

Refs.~\cite{bishop2020effect,bishop2022effect} showed, recently, that a shell composed of viscous fluid surrounding a GW event modifies the magnitude of the GWs according to a formula that reduces to Eq.~\eqref{eqn:eta} when the matter is far from the GW source, but can be much larger when the matter is at a distance comparable to the wavelength $\lambda$. Energy loss by GW damping must be balanced by energy gain to the damping medium, and Ref.~\cite{kakkat2024gravitational} calculated the temperature increase in a spherically symmetric viscous shell surrounding a circular binary source. One of the results of this heating may be the possible emission of electromagnetic (EM) waves as suggested by ~\cite{milosavljevic2005afterglow,tanaka2010time,li2012gravitational}. For a stationary accretion disk, modelled on data from the binary black hole merger GW 150914, it was found that the temperature rise inside the viscous shell was approximately $10^7$K. 

GW heating and damping is expected to be significant when the wavelength $\lambda$ is somewhat larger than the distance $r$ between the viscous medium and the GW source~\cite{Bishop2024essay}. In this paper, we discuss various astrophysical and cosmological scenarios to which these conditions apply: (a) We consider both the damping and heating effects following a binary neutron star (BNS) merger; (b) GW damping of core collapse supernae (CCSNe) was previously investigated~\cite{bishop2022effect}, and here we consider the heating effect; (c) we investigate the damping and heating effects for primordial GWs generated by electroweak first order phase transitions (EW FOPTs).

This paper is organized as follows: In the first section (\ref{sec:previous}), for completeness, we summarize previous work and the expressions obtained for GW damping and heating in a viscous shell surrounding a GW source. 
In section \ref{sec:astro} we apply our analysis to the various astrophysical and cosmological scenarios described above.
Finally, in section \ref{sec:summary}, we provide a summary and conclusion of our findings.

\section{Previous work}\label{sec:previous}

The Bondi-Sachs formalism is a well-known mathematical framework used in general relativity \cite{bishop1999incorporation,bishop1997high,gomez2001gravitational}. 
Consider the Bondi-Sachs metric representing a general spacetime \cite{sachs1962gravitational,bondi1962gravitational} in null coordinates $(u,r,x^A)$ as
\begin{align}
    ds^2=&-\left(e ^{2\beta}(1+W_{c}r)-r^2h_{AB}U^AU^B\right)du^2
    -2e^{2\beta}dudr\nonumber\\
    &-r^2h_{AB}U^B dudx^A+r^2h_{AB}dx^Adx^B,
\end{align}
where $h_{AB}h^{BC}=\delta_C ^A$, $\det(h_{AB})=\det(q_{AB})$, $q_{AB}$ is the canonical metric on the unit sphere. Here coordinate $u$ labels the null outgoing hypersurface, $r$ coordinate is the surface area coordinate, and $x^A=(\theta,\phi)$ are spherical polar coordinates. As shown in previous work, $h_{AB}$ can be represented by a spin-weight 2 complex qunatity $J$, and $U^A$ can be represented by a spin-weight complex quantity $U$; see \cite{gomez1997eth,bishop1999incorporation,newman1966note} for a more detailed explanation.

	We make the ansatz of small quadrupolar perturbations about Minkowski spacetime with the metric quantities $\beta,U,W_c,J$ taking the form
	\begin{align}
		\beta=&\Re(\beta^{[2,2]}(r)e^{i\nu u}){}_0Z_{2,2}\,,\;\;
		U=\Re(U^{[2,2]}(r)e^{i\nu u}){}_1Z_{2,2}\,,\nonumber \\
		W_c=&\Re(W_c^{[2,2]}(r)e^{i\nu u}){}_0Z_{2,2}\,,\;\;
		J=\Re(J^{[2,2]}(r)e^{i\nu u}){}_2Z_{2,2}\,.
		\label{e-ansatz}
	\end{align}
	The perturbations oscillate in time with frequency $\nu/(2\pi)$. The quantities ${}_s Z_{\ell,m}$ are spin-weighted spherical harmonic basis functions related to the usual ${}_s Y_{\ell,m}$ as specified in~\cite{bishop2005linearized,bishop2016extraction}. They have the property that ${}_0 Z_{\ell,m}$ are real, enabling the description of metric quantities without mode-mixing.
	
Substituting the anzatz Eq.~\eqref{e-ansatz} into the linearized vacuum Einstein equations leads to a set of coupled ordinary differential equations for $\beta^{[2,2]}(r),W_c^{[2,2]}(r),J^{[2,2]}(r),U^{[2,2]}(r)$ whose solution (in the case of purely outgoing perturbations) comprises polynomials in $1/r$ ~\cite{bishop2005linearized}; these solutions are given explicitly in our previous paper~\cite{kakkat2024gravitational} and are not repeated here.

We now consider the case that the GW source is surrounded by a shell of matter. Due to the GW perturbations, the matter within the shell undergoes motion, and the velocity field is calculated using the matter conservation conditions~\cite{bishop2022effect}. Having found the velocity field, it is then straightforward to calculate the shear tensor $\sigma_{ab}$, and explicit formulas are presented in~\cite{bishop2022effect}.
It is shown in \cite{baumgarte2010numerical} that
\begin{equation}\label{eqn:Et}
    \frac{\partial_u E_{shell}}{\Delta V}=2\eta\sigma_{ab}\sigma^{ab}\,,
\end{equation}
where $E_{shell}$ is the energy in an element of the shell with volume $\Delta V$, and $\eta$ is the coefficient of (dynamic) viscosity. Then, using Eq.~(18) in~\cite{bishop2022effect}, it follows that GWs are damped according to the formula 
\begin{equation}
H(r_o)=H(r_i)\exp\left(-\frac{8\pi G\eta}{c^3}(f(r_o)-f(r_i))\right)\,,
\label{e-Hro}
\end{equation}
where conversion to SI units has been made, with $G$ the gravitational constant and $c$ the speed of light; $H$ denotes the magnitude of the GWs rescaled to account for a $1/r$ falloff, and $r_o$ and $r_i$ represent the outer and inner radii of the shell, respectively; and
\begin{equation}
f(r)=r-\frac{\lambda^2}{2r\pi^2}-\frac{3\lambda^4}{r^32^4\pi^4}
-\frac{9\lambda^6}{r^52^6\pi^6}-\frac{45\lambda^8}{r^72^8\pi^8}\,,
\end{equation}
with $\lambda$ being the wavelength of the GWs.

In~\cite{kakkat2024gravitational} we obtained an expression for the energy input to the shell as a function of the angular coordinates. However, for the applications to be considered here, the angular dependence of the GWs is uncertain, so we treat the heating effect as uniform across the shell. Then the $Y_{0,0}$ terms in the formulas in~\cite{kakkat2024gravitational} are retained, but those terms with higher order spherical harmonics ($Y_{2,0},Y_{4,0}$) are discarded. Eq.~(15) of~\cite{kakkat2024gravitational} simplifies to
\begin{equation}\label{eqn:eshell}
    \frac{\partial_u E_{shell}}{\Delta V}=\frac{\sqrt{\pi}}{6}\nu^2\eta \partial_u E_{GW}D_0\,,
\end{equation}
where $\partial_u E_{GW}$ is the power output of GWs, and where
\begin{equation}
D_0= \frac{12({\nu}^8r^8+2{\nu}^6r^6+9{\nu}^4r^4+45{\nu}^2r^2+315)}{\sqrt{\pi}{\nu}^{10}r^{10}}.
\label{e-TA-D}
\end{equation}
Then the temperature increase in the shell is
\begin{equation}
T-T_0=\frac{\sqrt{\pi} G\eta}{6c^5 C\rho}\nu^2  \Delta E_{GW}D_0\,,
\label{eqn:diffT}
\end{equation}
where conversion to SI units has been made, and where $\rho$ is the density, $C$ is the specific heat capacity (in J/${}^\circ$K/kg), and $\Delta E_{GW}$ is the emitted GW energy. In the formula for $D_0$ in Eq.~\eqref{e-TA-D}, $r\nu\rightarrow r \nu/c=r/(2\pi\lambda)$.

\section{Astrophysical and Cosmological Scenarios for Gravitational Wave Heating}\label{sec:astro}

\subsection{The binary neutron star merger}
This section examines the GW damping and potential GW heating effect in the binary neutron star (BNS) merger case which is a source of GWs for LIGO/Advanced LIGO  and other ground-based detectors.
\subsubsection{Overview}
The first detection of GWs from a BNS merger occurred on August 17, 2017, labeled as GW170817;
a subsequent detection event was GW190425,
and further detections are expected. The merging process of BNS results in the ejection of a substantial amount of matter, and these ejected materials are recognized as active sites for nucleosynthesis, specifically the rapid neutron capture process (r-process), leading to the formation of heavy metals. Observations such as Killonova or Macronova provide direct evidence of electromagnetic emissions powered by the decay of heavy metals produced through the r-process~\cite{li1998transient,metzger2010electromagnetic,roberts2011electromagnetic,barnes2013effect,tanaka2013radiative}. A notable example in this context is the observation of ultraviolet, optical, and infrared signals associated with GW 170817, see details {in~\cite{abbott2017multi,tanaka2017kilonova,arcavi2017optical,coulter2017swope,cowperthwaite2017electromagnetic,drout2017light,evans2017swift,kasliwal2017illuminating,smartt2017kilonova,tanvir2017emergence}.}

The expulsion of a significant amount of matter during the merger of two neutron stars is a well-established phenomenon \cite{rosswog1998mass,ruffert1996coalescing,freiburghaus1999r}. This ejected nuclear matter typically coalesces into a disk/ torus shape around the system. For a detailed analysis of the evolutionary process and the formation, see \cite{fernandez2013delayed,metzger2014red,perego2014neutrino,fernandez2015outflows,just2015comprehensive,siegel2017three,fujibayashi2018mass,fernandez2019long,baumgarte1999maximum} and the references therein.

To understand the mechanism and process in the evolution of merging of BNS, considerable efforts have been given to numerical simulations over the years.  A well-known finding from simulations~\cite{demorest2010two,antoniadis2013massive,shibata2005merger,shibata2006merger,kiuchi2009long,hotokezaka2011binary,hotokezaka2013remnant,takami2015spectral,dietrich2015numerical,bernuzzi2016loud,ciolfi2017general} indicates that, for a remnant mass of $m\lesssim 2.8 M_\odot$, the remnant, albeit temporarily, transforms into a hypermassive neutron star
(HMNS) rather than immediately collapsing into a black hole, regardless of the equation of state (EOS) employed. 
Moreover, when the remnant mass exceeds $2.8 M_\odot$, a black hole could be formed, and the mass of the black hole is heavily influenced by the EOS.

\subsubsection{GW heating in binary neutron star merger}

It has been shown recently that when gravitational waves traverse through the shell of matter enveloping the source, energy is transferred to the shell, leading to an increase in temperature inside the shell \cite{kakkat2024gravitational}. To quantify this temperature rise in the BNS merger case, we employ numerical simulation models as described in~\cite[Table I]{bernuzzi2016loud}. The models outlined in \cite{bernuzzi2016loud} are specifically configured based on observed BNS systems. Furthermore, it has been demonstrated in~\cite{bernuzzi2016loud} that the GW energy emitted within the initial $\sim 10$ms of the resulting HMNS is approximately twice the energy emitted throughout the entire inspiral history of the binary. Additionally, approximately $0.8-2.5\%$ of the binary mass-energy is emitted at kilohertz (kHz) frequencies during the early evolution of the HMNS.

In the case of the postmerger signal, the frequency $f$ varies over the range $1-2$kHz, and the radius $r$ is set to be $12$km. The density is approximated as $10^{16}$kg/m${}^3$. The value of specific heat is not very well known in the literature. However, \cite{cumming2017lower} provides an estimation for the specific heat $C$. Assuming the neutron star
KS1731-260 has mass $1.4M_\odot$, radius $12~km$, and temperature $10^8~K$, then
its heat capacity is computed using \cite[Eqn. 12]{cumming2017lower} as $5.84$J/kg/K. Over the models specified in \cite[Table I]{bernuzzi2016loud}, the value $\Delta E_{GW}$ varies over $1.86\times 10^{44}$ to $2.6\times 10^{45}$J. Furthermore, the reported values of viscosity span from $10^{24}$ to $10^{30}$ J sec/m³~\cite{janka1995can,kochanek1992coalescing,bildsten1992tidal}. A summary of parameter values is provided in Table \ref{parameter1}.

 \begin{table}[th]
 \begin{tabular}{ll}
 \hline
	Parameter & Range   
	\\
	\hline
	\hline
    Radius & $ 12000$ m 
   \\
   Frequency & 1000 - 2000 Hz
\\
   Density & $10^{16}$ kg/m³ 
   \\
   GW energy, $\Delta E_{GW}$ & $1.86\times 10^{44} -2.6\times 10^{45}$ J 
   \\
   Viscosity ($\eta$) & $10^{24} - 10^{30} $ \,kg/m/s 
   \\
   Specific heat & 5.84 J/kg/K
   \\
	\hline
	\hline
	\end{tabular}
	\caption{Parameter range for post-BNS merger case }
	\label{parameter1}
	\end{table}

 Employing parameters specified in table \ref{parameter1}, equation~\eqref{eqn:diffT} produces a temperature difference 
\begin{equation}
8.7 \times 10^{12}\,\mbox{K}\le (T-T_0)\le 3.0\times 10^{22}\,\mbox{K}\,.
\end{equation}

\subsubsection{GW Damping in BNS merger}

Previous studies have demonstrated that under certain conditions, the damping of GWs as they propagate through a viscous fluid can be notably significant. This is particularly evident in the cases of GWs originating from the CCSNe and the primordial GWs generated during the inflationary period in the early Universe \cite{bishop2022effect}. In this section, we illustrate the damping effects of GWs from the BNS merger source surrounded by a viscous shell. The parameters are outlined in Table \ref{parameter1}. The frequency ranges from $1000$ to $2000$ kHz, viscosity from $10^{24}$ to $10^{30}$ kg/m/s, and the inner and outer shell radii are $r_i=12$km and $r_o=24$km. Substituting these values into Eq.~\ref{e-Hro} yields a damping factor
\begin{equation}
0<\frac{H(r_o)}{H(r_i)}<0.98\,,
\end{equation}
so that, depending on the physical parameters, the damping varies between being complete to having only a minor effect. 

\subsubsection{Discussion}
\label{BNS-D}
The results derived above use a model of small perturbations about a Minkowski background, which conditions do not apply to a BNS even post-merger. Further, the modeling takes no account of physical processes that would dissipate the temperature increase. Thus the results should not be interpreted as quantitative predictions, but rather as qualitative, that damping of the post-merger GW signal may be highly significant and that the energy in the GWs may be transferred as heat to the region around the HMNS, so affecting its physics. 
It is shown in~\cite{flowers1979transport} that when the temperature rise is ${T> 10^9}$K, the neutron-neutron interaction dominates over the other interactions, and in fact, the temperature rise inside the viscous shell around the BNS merger is above the superfluid transition temperature. A similar result concerning GW170817 can be found in \cite{orsaria2019phase}. 

To date, there has been no observation of a BNS with a post-merger GW signal. While that is consistent with the results above, it is also consistent with the possibility of the prompt formation of a black hole, for which the quasinormal mode frequency would be outside the LIGO detection band. However, if such a signal is observed in future, our results could be used to impose an upper limit on the effective viscosity.

\subsection{Core-collapse supernovae}
Core-collapse supernovae (CCSNe) are a promising sources of GWs~\cite{Cutler:2002me}. However, all the GW events detected to date are of compact object mergers (COMs). The detection rate of galactic supernovae (SN) has been calculated to be about $1.63 \pm 0.46$ per century \cite{Rozwadowska:2020nab}, and with SN1987A the last detected supernova in our galaxy, the next galactic SN  event is already heading our way with detection imminent. With veteran GW detectors and neutrino detectors lying in wait, together with a host of electro-magnetic (EM) detectors, the next detectable CCSNe is expected to be the most significant event in multi-messenger astronomy (MMA)~\cite{Fryer:2023ehc,DiPalma:2023hxs}. The event  is expected to provide a wealth of information on the composition and the EoS of NSs and on the CCSNe evolution itself.

\subsubsection{Evolution of CCSNe}
There exist several reviews of CCSNe and the associated processes and phenomena ~\cite{Muller:2020ard,Abdikamalov:2020jzn,Heger03,Woosley02,Woosley06,Woosley:2006ie}. EM detection of SN only provide limited information on the interior regions of the star as photons originate at the outer edge. GW's originate as a consequence of the aspherical motion of the inner regions of the star so will provide greater clues to these regions as well as the mechanism leading to the SN explosion.
Multi-dimensional simulations have been carried out for GW's expected to be generated from CCSNe
~\cite{Andresen:2016pdt,Andresen:2018aom,Radice:2018usf}. These  are all computationally demanding with uncertainty still remaining on many of the parameters especially those within the inner regions of the stars. 

Core collapse occurs on a timescale of the order of $ {\sim} 0.3\,\mathrm{s}$, with the core splitting into an outer part, which plunges supersonically, and an inner core, collapsing at subsonic speed. The collapse of the inner core will stop when supranuclear densities are reached (where the nuclear matter stiffens),  resulting in a  bounce of the inner core. This causes a shock wave to be launched into the collapsing outer core. The shock loses energy to dissociation of iron nuclei, and this in turn leads to the stalling of the shock wave around ${\sim} 150\,\mathrm{km}$ within ${\sim} 10\,\mathrm{ms}$ after formation. Once the shock wave has stalled the infalling shells of stellar matter will pile on the proto neutron star (PNS). Unless there is a \textit{revival} of the shock within a few hundred milliseconds, the PNS collapses directly to a BH. In the case of a \textit{revival}, the resulting outward shock expels the infalling outer shells, leaving behind a stable NS. For the \textit{revival} to occur there needs to be sufficient energy and we investigate whether the heating effect, through gravitational damping, could be sufficient to provide this, hence preventing the PNS collapsing to a BH and, instead, leading to a NS. This would have further implications for current population synthesis of NS's from CCSNe. In addition, further GW's would be generated as a result of the \textit{revival}.

\subsubsection{Calculations for GW heating for CCSNe}
The detectable frequencies of GWs expected from CCSNe range from 100 to 1300 Hz~\cite{hsieh2023new}. Estimates for the radius of the core of the PNS range from 10 to 30 km.
A summary of various investigations is given in Table~\ref{parameter}.
	\begin{table}[th]
	\begin{tabular}{lll}
	\hline
	Reference  &$r_i$ [m] &Frequency [Hz]\\
	\hline
	\hline
	Roma~\cite{Roma:2019kcd}	&& 96 - 1000 \\
	Scheidegger~\cite{Scheidegger:2010en}	&&317 - 935 \\
	M\"{u}ller~\cite{Muller:2011yi}	&&130 - 1100 \\
	Andresen~\cite{Andresen:2016pdt}
	&$10^{4} - 2.8\times 10^{4}$&100 - 700\\
	Kuroda~\cite{Kuroda:2016bjd} &$10^{4} - 2\times 10^{4}$& 100 - 671\\
	Mezzacappa~\cite{Mezzacappa:2020lsn}	&$10^{4} - 2\times 10^{4}$& 200 - 600\\
	Powell~\cite{Powell:2018isq}	  &$10^{4}$&800 - 1000\\
	Shibagaki~\cite{Shibagaki:2020ksk}	
	&$10^{4}$&200 - 800\\
	\hline
	\hline
	\end{tabular}
	\caption{Parameter values for CCSNe from various references for the core radius and GW frequency }
	\label{parameter}
	\end{table}

The GW energy released from a CCSNe is expected to be approximately $10^{43} - 10^{46}$ erg~\cite{Abdikamalov:2020jzn}.
The density at core bounce is expected to be approximately $3.2 \times 10^{14}$ kg/m³~\cite{Furusawa:2022ktu}. 
The viscosity in the core collapse environment receives contributions from the neutrino viscosity, the turbulent viscosity caused by the magnetorotational instability (MRI), and the turbulent viscosity by entropy and composition-gradient-driven convection. ~\cite{Thompson:2004if} find that the MRI will operate and dominate the viscosity within the PNS even for the slowest rotators considered, dominating the neutrino viscosity by 2 to 3 orders of magnitude. 	
In the region of the shell, say between 10 and 30\,km, the kinematic viscosity $\gamma$ varies in the range $10^8$ to $10^{10}$m${}^2$/s. 
Multiplication by the density $\rho$, about $10^{15}$kg/m${}^3$, 
gives the dynamic viscosity $\eta$ at $10^{23}$ to $10^{25}$\,kg/m/s.  Ref.~\cite{Spruit:2001tz} found values consistent with these magnitudes. Values for $\eta$ in neutron star material are discussed in~\cite{Kolomeitsev15}  and can be as high as $10^{22}$\,kg/m/s.
For the specific heat (C), we would expect this to be no less than the value taken for the BNS case and so take  the value of 6 J/kg/K as a lower limit.
We summarise the the range for the parameters of CCSNe in
Table~\ref{parameter2}.
 \begin{table}[th]
 \begin{tabular}{ll}
 \hline
	Parameter & Range   
	\\
	\hline
	\hline
   Core radius & $10^{4} - 3\times 10^{4}$ m 
   \\
   Frequency & 100 - 1000 Hz
\\
   Density & $10^{14} - 10^{16}$ kg/m³ 
   \\
   GW energy, $\Delta E_{GW}$ & $10^{36} - 10^{39}$ J 
   \\
   Kinematic viscosity ($\gamma$) & $10^8 - 10^{10}$\, m${}^2$/s
   \\ 
   Dynamic viscosity ($\eta$) & $10^{23} - 10^{25} $ \,kg/m/s 
   \\
   Specific heat & 6 - 200 J/kg/K
   \\
	\hline
	\hline
	\end{tabular}
	\caption{Parameter range for CCSNe}
	\label{parameter2}
	\end{table}


We now use Eq.~\eqref{eqn:diffT}, with 
$D_{0}$ from Eqs.~\eqref{e-TA-D},  
and the values in Table~\ref{parameter2} to calculate the range of the expected temperature increase. Note that although values of $\rho,\eta$ are listed in the Table, they are not used in the calculation; rather, we use only $\gamma$ since replacing $\eta$ by $\rho\gamma$ in Eq.~\eqref{eqn:diffT} leads to the cancellation of $\rho$. We find
\begin{equation}
38\,\mbox{K}\le (T-T_0)\le 7.0\times 10^{20}\,\mbox{K}\,,
\label{e-CSSNe-dT}
\end{equation}
so that, depending on the actual values of physical paramaters, the significance of the effect varies from very high to none.

\subsubsection{Discussion}
Noting that the comments made in Section~\ref{BNS-D} about the applicability of the model also apply here, we now consider the case that the heating effect is towards the top of the range in Eq.~\ref{e-CSSNe-dT}.

Not all the energy may contribute to an increase in temperature, and there may be contributions to other processes such as phase transition. Conversely, the heating effect may, itself, contribute to phase transition. This may be important when the revival at core bounce is insufficient to produce a SN. 
~\cite{Sagert2009,Fischer_2018} demonstrated that a first-order phase transition from nuclear matter to a quark-gluon plasma can also trigger a SN explosion if other mechanisms do not succeed. When phase transition sets in, it instigates a dynamical collapse of the PNS to a more compact hybrid star with a core of pure quark-gluon plasma. A bounce shock is then formed when the quark core settles into a new hydrostatic equilibrium. ~\cite{Sagert2009,Fischer_2018} showed that this second shock can revive the first shock that is linked to the instant of core bounce. Thus a failed SN may be rescued by this second shock if it is powerful enough to launch a SN explosion, with the energy needed for the explosion released at the expense of gravitational binding energy.

\subsection{Primordial gravitational waves.}

\subsubsection{Overview}

Possible sources of GWs in the early epochs of the Universe include: inflation according to the standard model (SM), mechanisms in inflation of beyond standard model (BSM) theories, preheating and similar non-perturbative phenomena, first order phase transitions (FOPTs) and topological defects. These generate a fundamentally stochastic GW background analogous to the cosmic microwave background in the electromagnetic spectrum. Of the FOPTs, certain BSM scenarios propose that the electroweak (EW) FOPT overlaps in frequency and magnitude with the detection sensitivities of future GW detectors. In particular, the detection design on the Laser Interferometer Space Antenna (LISA) makes provision for detecting EW FOPTs \cite{Caprini2020}. For comprehensive reviews of cosmological GWs see, \cite{Caprini2018} and \cite{Auclair2023}.

The EW FOPT is a symmetry breaking process of nucleation where a scalar field, which can be a Higgs or Higgs-like field, is percolated as bubbles in the EW plasma consisting of SM particles. Whereas the SM predicts a smooth cross over between the EW epoch and the subsequent quark-gluon epoch, thermal FOPTs proposed by some BSMs can be rapid and highly energetic events where bubbles expand and collide to combine as larger bubbles, which interacts with the surrounding plasma to generate sound waves and turbulence. The collision of bubbles itself, in the scalar field, along with both the propagating sound waves and turbulence form separate sources of GWs, with the sound wave propagation being the dominating effect. More detailed explanations of the EW FOPT and its detection are presented in  \cite{Weir2018} and \cite{Ramsey-Musolf2020}.

Detection of the EW FOPT, will be of considerable interest to proponents of BSMs while the non-detection will substantiate the cross over anticipated by the SM predictions. There can, however, be other reasons for non-detection that still fall within BSM theories, or the interference of other physical processes, such as viscous damping. With the latter in mind, we now consider scenarios where the linearized model, presented in previous sections, is applied to the EW FOPT process. Our approach here is to consider a discrete source generating GWs, which are propagated in the quark-gluon plasma following the EW phase transition. The linearized model applies to this in the sense that the plasma in the quark-gluon epoch is considered to be a dense shell through which the GWs propagating from the EW phase transition are travelling. This is a naive first approximation to understand the viscous effects on the EW FOPT GWs that will provide guidance for planned research where the effects of the full radiation background will be studied. See \cite{MironGranese2021} for a detailed study based on physical processes of viscous effects on primordial GWs.

\subsubsection{Damping}

The reasoning we follow now, is that after the EW phase transition, the GWs propagate through the quark-gluon plasma present in the quark-gluon epoch. From \cite{Schafer2009}, we use the viscosity as $\eta = 5 \times 10^{11}$ kg/m/s, which when multiplied by $G/c^3$ to convert to geometric units is: $\eta = 1.235 \times 10^{-24}$ m$^{-1}$. We get the density from the kinematic viscosity stated in \cite{Trachenko2021} as $\nu = \mu/\rho = 10^{-7} \Rightarrow \rho = \mu/\nu \sim 10^{18}$ kg/m$^{3}$.

We consider the expected measurable frequencies for GWs generated from the electroweak transition to the quark-gluon epoch. From \cite{Caprini2020}, we take a representative peak observable frequency as $\sim 10^{-2}$ Hz. We then apply the same reasoning as that in \cite{bishop2022effect} with the time and temperature of the EW FOPT as $t=10^{-12}$ s and $T=10^{15}$ K, respectively. According to \cite{Misner1973-mq} (Section 28.1), $T$ is proportional to $1/a(t)$. Thus, at the end of the EW FOPT, we take $a\sim 10^{-15}$. From this it follows that the peak frequency at the time of generation is $f_{peak} \approx 10^{12}$ Hz, which amounts to a wavelength $\lambda_{peak} \approx 3 \times 10^{-4}$ m.

Using $\mathcal{H} = \dot{a}/a$ \footnote{The Hubble rate is denoted by $\mathcal{H}$ instead of the usual $H$ to avoid confusion with the metric perturbation, defined earlier} with $a(t) \propto t^{1/2}$, for a radiation Einstein-de Sitter model, we take $\mathcal{H}=1/2\cdot t^{-1} = 0.5 \times 10^{12}$. From $c/\mathcal{H}$, it then follows that the Hubble radius is $R_h = 6 \times 10^{-4}$ m.

In \cite{bishop2022effect}, it was shown that the GW damping effect is given by
\begin{align}
\frac{dH}{dr}= & -8\pi\eta H
	\left(1+\frac{2}{r^2\nu^2}+\frac{9}{r^4\nu^4} \right.  \nonumber \\
	& \left. +\frac{45}{r^6\nu^6}+\frac{315}{r^8\nu^8} \;
	\right)\,, \label{eqn:dHdr}
\end{align}
where $r$ is the distance form the source and $\nu/(2 \pi )$ is the GW frequency. With the $R_h$ and $\lambda$ being of similar order, i.e., $10^{-4}$ m, neither of the simplifying assumptions of $\lambda \ll r_i$ nor $\lambda \gg r_i$, presented in \cite{bishop2022effect} are applicable. It is therefore, necessary to solve \eqref{eqn:dHdr} without simplification. For the cosmological scenario, we rewrite \eqref{eqn:dHdr} in terms of $t$ by introducing
\begin{equation}
r_{\nu 2} = \frac{\lambda^2 \, t}{t_i \left( 2 \pi(r_i c t-c t_i) \right)} \label{eqn:rnu2} \;.
\end{equation}
Substituting \eqref{eqn:rnu2} into \eqref{eqn:dHdr} with some manipulation then gives
\begin{align}
\frac{d \log H}{dt} &= - \frac{8\pi\eta G}{ct} \left( 1 + 2 r_{\nu 2} + 9 {r_{\nu 2}}^2 \right. \nonumber \\
    & \left.  + 45 {r_{\nu 2}}^3 + 315 {r_{\nu 2}}^4 \;
    \right) \; . \label{eqn:dHdt}
\end{align}
Equation \eqref{eqn:dHdt} is solved using computer algebra and then the factor of damping is evaluated for different radii starting with $R_h$ as maximum \footnote{The form in which \eqref{eqn:dHdt} is presented here is favourable for integration with the computer algebra system \emph{Maple}.}. It is found that damping is negligible on the larger range of radii and more prominent on smaller scales. This is an effect that exhibit a sudden transition from no damping at $r_i = 10^{-7}$ to sizeable damping at  $r_i = 10^{-8}$, with a damping factor of $0.023$, to complete damping at $r_i = 10^{-9}$.

\subsubsection{Heating}

In this section, we experiment with the linearized equations along with cosmological and nucleation parameters to establish some baseline values for heating following the EW FOPT. These are mostly rudimentary assumptions towards gaining insight into processes that affect the detection and physics of GWs in the early Universe.

\begin{figure}[h!]
	\begin{center}
    	\includegraphics[width=0.95\columnwidth,angle=0]{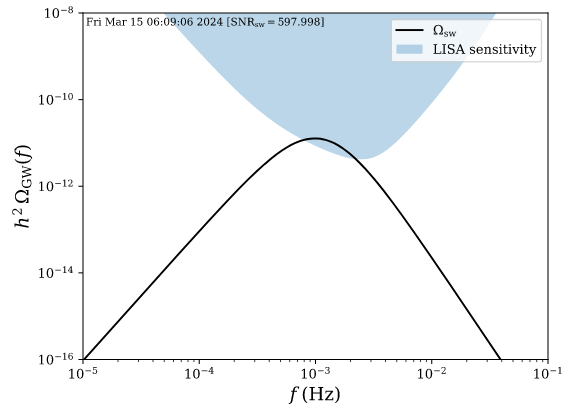}
    	\caption{Using the POPT online tool, described in \cite{Caprini2020} with the parameter values $v_w=0.9$, $\beta/\mathcal{H}_{*}=50$, $\alpha_{\theta}=0.2$, $T_{*}=200$GeV, $g_{*}=100$. }
    	\label{fig:bubble-parameters}
	\end{center}
\end{figure}

For the purposes of representing FOPT detection by LISA, formulae that are model agnostic, to some extent, are used to represent the physics of the nucleation process \cite{Weir2018, Auclair2023, Caprini2020, Renzini2022}. These formulae make use of the parameters in Table \ref{cos:bubble-parameters} to represent the interaction of the nucleating bubbles.
\begin{table}[th]
 \begin{tabular}{ l p{0.7\linewidth}}
 \hline
	Parameter & Description
	\\
	\hline
	\hline
    $\beta^{-1}$ & Phase transition duration. \\
    $R_*$        & Size of the bubble towards the end of the PT. \\
    $v_w$        & Bubble wall speed. \\
    $\mathcal{H}_*^{-1}$   & Hubble time at the time of PT. \\
    $g_{*}$      & Degrees of freedom. \\
    $\alpha$     &  The GW energy budget ratio GWs generation vs. radiation ($\rho_{vac}/\rho_{rad}$) \\
	\hline
	\hline
	\end{tabular}
	\caption{Parameters used in nucleation formulaes for the LISA design \cite{Caprini2020}. }
	\label{cos:bubble-parameters}
	\end{table}
These parameters are then used to determine possible intersections between the LISA sensitivity curve and the GW frequency spectrum, which should then allow distinction between candidate models when observations become available. Typical values for these parameters are shown in Figure \ref{fig:bubble-parameters}.

Table \ref{cos:parameters} lists the values required in the linear models, presented in the previous section, with the nucleation parameters along values already determined and that will follow in this section.

For the specific heat capacity $(C)$, we make use of the lowest value presented in the previous sections since lower values will provide an upper bound to the heating effect. It is evident from \eqref{eqn:diffT}, that increasing the value of $C$ by an order will have the inverse effect on the order of $\Delta T$.

As a radius value, the bubble radius is calculated using
\begin{equation}
R_{*} = (8\pi)^{1/3} \frac{v_{w}}{\beta} \label{eqn:R_star}
\end{equation}
with $\beta = 50 \, \mathcal{H}_{*}$, taking $\mathcal{H}_{*}=0.5 \times 10^{12}$ s$^{-1}$ as determined earlier, it follows that $R_{*} = 1.582 \times 10^{-5}$ m. As expected, the maximum bubble radius is somewhat smaller than the Hubble radius, determined earlier. The Hubble radius is therefore a better value as an extreme upper bound while not a complete over estimation.

To get a measure of energy, we consider that some portion of the total mass of the Hubble sphere is converted into energy generating GWs ($\Delta E_{GW}$). The mass of the Hubble sphere is calculated as
\begin{equation}
M_{R_h} = \frac{4}{3} \pi {R_h}^2  = 4.524 \times 10^{9} \text{kg} \label{eqn:Mrh}
\end{equation}
using $R_h = 6 \times 10^{-4}$ m, as previously determined. Using $E=mc^2$, this translates to a rest mass energy of $E_{R_h} = 4.072 \times 10^{26}$ J .
 \begin{table}[th]
 \begin{tabular}{ l p{0.5\linewidth}}
 \hline
	Parameter & Range
	\\
	\hline
	\hline
    Bubble radius ($R_{*}$) & $1.5818 \times 10^{-5}$ m
   \\
   Frequency peak ($f_{peak}$) &  $10^{12}$ Hz
\\
   Density ($\rho$) & $\sim 10^{18}$ kg/m³
   \\
   GW energy, ($\Delta E_{GW}$) & $< 10^{26} - 10^{30}$ J
   \\
   Viscosity ($\eta$) & $5 \times 10^{11} $ \,kg/m/s
   \\
   Specific heat ($C$) & 5.84 J/kg/K
   \\
	\hline
	\hline
	\end{tabular}
	\caption{Parameter range for GWs generated by the EW FOPT and propagating through a quark-gluon plasma.}
	\label{cos:parameters}
	\end{table}

As an extreme upper bound where the mass of the Hubble sphere is completely converted into energy, i.e., $\Delta E_{GW} = E_{R_h}$, substituting the values in Table \ref{cos:parameters} into Eqs. \eqref{e-TA-D} and \eqref{eqn:diffT} yields the temperature increase as
\begin{equation}
\Delta T \sim 10^{-10} \, \mathrm{K} \; .
\end{equation}
Alternatively, using a Hubble radius derived from the $\Lambda$CDM model, $R_h = 0.013m$, see \cite{Melia2022}, the corresponding energy follows as $\Delta E_{GW} \sim 10^{30}$J with the temperature increase resulting in
\begin{equation}
\Delta T \sim 10^{-6} \, \mathrm{K} \; .
\end{equation}
In both cases the expected temperature increase is yielded as insignificant, given that a small portion of the mass is realistically expected to be converted to GW energy.

\subsubsection{Discussion}
For primordial GWs, the linearized approach followed here, aligns with the expectation that GWs generated for an EW FOPT will be detectable without deformation from damping for GWs of the scales of the expected bubble radii used in the LISA design. It was, however, found that on smaller scales $r_i \sim 10^{-7}$ to $10^{-9}$, there is a rapid transition from no damping to GWs being fully damped. Thus  a future non-detection of GWs from EW FOPTs could mean that dominating bubble radii are substantially smaller than expected.

Extending the calculations to the temperature increase due to EW FOPT GW energy transfer when GWs propagate through the quark-gluon epoch, the linearized approach suggests that the effect will be negligible. This follows from calculating an extreme upper bound where the mass of the complete Hubble sphere is converted to GWs but still resulted in a negligible increase in temperature.

\section{Summary and conclusions}\label{sec:summary}

In this article, we have applied our previously developed theory of GW damping and heating to various scenarios in which the effects are expected to be important. There are feasible values of the astrophsyical parameters which lead to complete damping of the post-merger GW signal from a BNS, and the resultant heating would be significant to the behavior of the remnant. Similarly, the heating effect on a CCSNe may be significant to the revival of the stalled shock wave, which is needed for the model to lead to a supernova explosion. GWs generated by EW FOPTs are expected to have little impact on the temperature of the surrounding medium, and are unlikely to be damped unless the dominating bubble radii are substantially smaller than current estimates.

The results obtained in this work model GWs as small perturbations on a Minkowski background, and treat the viscous medium as a static viscous shell, and this means that there are unknown limitations on the accuracy of our results. Further, the uncertainty in the values of various astrophysical parameters leads to our predictions covering a known, but large, range. Nevertheless, it is clear that GW damping and heating can be significant, and that these effects should be included in modelling the processes considered here.

 \begin{acknowledgements}

VK and ASK express their sincere gratitude to Unisa for the Postdoctoral grant and their generous support. We would also like to thank Nahuel Mir\'{o}n-Granese for useful suggestions on the cosmological section.
\end{acknowledgements}

\bibliography{vis}

\begin{thebibliography}{104}%
\makeatletter
\providecommand \@ifxundefined [1]{%
 \@ifx{#1\undefined}
}%
\providecommand \@ifnum [1]{%
 \ifnum #1\expandafter \@firstoftwo
 \else \expandafter \@secondoftwo
 \fi
}%
\providecommand \@ifx [1]{%
 \ifx #1\expandafter \@firstoftwo
 \else \expandafter \@secondoftwo
 \fi
}%
\providecommand \natexlab [1]{#1}%
\providecommand \enquote  [1]{``#1''}%
\providecommand \bibnamefont  [1]{#1}%
\providecommand \bibfnamefont [1]{#1}%
\providecommand \citenamefont [1]{#1}%
\providecommand \href@noop [0]{\@secondoftwo}%
\providecommand \href [0]{\begingroup \@sanitize@url \@href}%
\providecommand \@href[1]{\@@startlink{#1}\@@href}%
\providecommand \@@href[1]{\endgroup#1\@@endlink}%
\providecommand \@sanitize@url [0]{\catcode `\\12\catcode `\$12\catcode
  `\&12\catcode `\#12\catcode `\^12\catcode `\_12\catcode `\%12\relax}%
\providecommand \@@startlink[1]{}%
\providecommand \@@endlink[0]{}%
\providecommand \url  [0]{\begingroup\@sanitize@url \@url }%
\providecommand \@url [1]{\endgroup\@href {#1}{\urlprefix }}%
\providecommand \urlprefix  [0]{URL }%
\providecommand \Eprint [0]{\href }%
\providecommand \doibase [0]{https://doi.org/}%
\providecommand \selectlanguage [0]{\@gobble}%
\providecommand \bibinfo  [0]{\@secondoftwo}%
\providecommand \bibfield  [0]{\@secondoftwo}%
\providecommand \translation [1]{[#1]}%
\providecommand \BibitemOpen [0]{}%
\providecommand \bibitemStop [0]{}%
\providecommand \bibitemNoStop [0]{.\EOS\space}%
\providecommand \EOS [0]{\spacefactor3000\relax}%
\providecommand \BibitemShut  [1]{\csname bibitem#1\endcsname}%
\let\auto@bib@innerbib\@empty
\bibitem [{\citenamefont {Hawking}(1966)}]{hawking1966perturbations}%
  \BibitemOpen
  \bibfield  {author} {\bibinfo {author} {\bibfnamefont {S.~W.}\ \bibnamefont
  {Hawking}},\ }\bibfield  {title} {\bibinfo {title} {Perturbations of an
  expanding universe},\ }\href@noop {} {\bibfield  {journal} {\bibinfo
  {journal} {Astrophysical Journal, vol. 145, p. 544}\ }\textbf {\bibinfo
  {volume} {145}},\ \bibinfo {pages} {544} (\bibinfo {year}
  {1966})}\BibitemShut {NoStop}%
\bibitem [{\citenamefont {Esposito}(1971)}]{esposito1971absorption}%
  \BibitemOpen
  \bibfield  {author} {\bibinfo {author} {\bibfnamefont {F.~P.}\ \bibnamefont
  {Esposito}},\ }\bibfield  {title} {\bibinfo {title} {Absorption of
  gravitational energy by a viscous compressible fluid},\ }\href@noop {}
  {\bibfield  {journal} {\bibinfo  {journal} {Astrophysical Journal, vol. 165,
  p. 165}\ }\textbf {\bibinfo {volume} {165}},\ \bibinfo {pages} {165}
  (\bibinfo {year} {1971})}\BibitemShut {NoStop}%
\bibitem [{\citenamefont {Marklund}\ \emph {et~al.}(2000)\citenamefont
  {Marklund}, \citenamefont {Brodin},\ and\ \citenamefont
  {Dunsby}}]{marklund2000radio}%
  \BibitemOpen
  \bibfield  {author} {\bibinfo {author} {\bibfnamefont {M.}~\bibnamefont
  {Marklund}}, \bibinfo {author} {\bibfnamefont {G.}~\bibnamefont {Brodin}},\
  and\ \bibinfo {author} {\bibfnamefont {P.~K.}\ \bibnamefont {Dunsby}},\
  }\bibfield  {title} {\bibinfo {title} {Radio wave emissions due to
  gravitational radiation},\ }\href@noop {} {\bibfield  {journal} {\bibinfo
  {journal} {The Astrophysical Journal}\ }\textbf {\bibinfo {volume} {536}},\
  \bibinfo {pages} {875} (\bibinfo {year} {2000})}\BibitemShut {NoStop}%
\bibitem [{\citenamefont {Brodin}\ \emph {et~al.}(2001)\citenamefont {Brodin},
  \citenamefont {Marklund},\ and\ \citenamefont {Servin}}]{brodin2001photon}%
  \BibitemOpen
  \bibfield  {author} {\bibinfo {author} {\bibfnamefont {G.}~\bibnamefont
  {Brodin}}, \bibinfo {author} {\bibfnamefont {M.}~\bibnamefont {Marklund}},\
  and\ \bibinfo {author} {\bibfnamefont {M.}~\bibnamefont {Servin}},\
  }\bibfield  {title} {\bibinfo {title} {Photon frequency conversion induced by
  gravitational radiation},\ }\href@noop {} {\bibfield  {journal} {\bibinfo
  {journal} {Physical Review D}\ }\textbf {\bibinfo {volume} {63}},\ \bibinfo
  {pages} {124003} (\bibinfo {year} {2001})}\BibitemShut {NoStop}%
\bibitem [{\citenamefont {Cuesta}(2002)}]{cuesta2002gravitational}%
  \BibitemOpen
  \bibfield  {author} {\bibinfo {author} {\bibfnamefont {H.~J.~M.}\
  \bibnamefont {Cuesta}},\ }\bibfield  {title} {\bibinfo {title}
  {Gravitational-to-electromagnetic wave conversion and gamma-ray bursts
  calorimetry: The grb980425/sn 1998bw~ 10 49 erg radio emission},\ }\href@noop
  {} {\bibfield  {journal} {\bibinfo  {journal} {Physical Review D}\ }\textbf
  {\bibinfo {volume} {65}},\ \bibinfo {pages} {064009} (\bibinfo {year}
  {2002})}\BibitemShut {NoStop}%
\bibitem [{\citenamefont {Bishop}\ \emph {et~al.}(2020)\citenamefont {Bishop},
  \citenamefont {van~der Walt},\ and\ \citenamefont
  {Naidoo}}]{bishop2020effect}%
  \BibitemOpen
  \bibfield  {author} {\bibinfo {author} {\bibfnamefont {N.~T.}\ \bibnamefont
  {Bishop}}, \bibinfo {author} {\bibfnamefont {P.~J.}\ \bibnamefont {van~der
  Walt}},\ and\ \bibinfo {author} {\bibfnamefont {M.}~\bibnamefont {Naidoo}},\
  }\bibfield  {title} {\bibinfo {title} {Effect of a low density dust shell on
  the propagation of gravitational waves},\ }\href@noop {} {\bibfield
  {journal} {\bibinfo  {journal} {General Relativity and Gravitation}\ }\textbf
  {\bibinfo {volume} {52}},\ \bibinfo {pages} {92} (\bibinfo {year}
  {2020})}\BibitemShut {NoStop}%
\bibitem [{\citenamefont {Bishop}\ \emph {et~al.}(2022)\citenamefont {Bishop},
  \citenamefont {van~der Walt},\ and\ \citenamefont
  {Naidoo}}]{bishop2022effect}%
  \BibitemOpen
  \bibfield  {author} {\bibinfo {author} {\bibfnamefont {N.~T.}\ \bibnamefont
  {Bishop}}, \bibinfo {author} {\bibfnamefont {P.~J.}\ \bibnamefont {van~der
  Walt}},\ and\ \bibinfo {author} {\bibfnamefont {M.}~\bibnamefont {Naidoo}},\
  }\bibfield  {title} {\bibinfo {title} {Effect of a viscous fluid shell on the
  propagation of gravitational waves},\ }\href@noop {} {\bibfield  {journal}
  {\bibinfo  {journal} {Physical Review D}\ }\textbf {\bibinfo {volume}
  {106}},\ \bibinfo {pages} {084018} (\bibinfo {year} {2022})}\BibitemShut
  {NoStop}%
\bibitem [{\citenamefont {Kakkat}\ \emph {et~al.}(2024)\citenamefont {Kakkat},
  \citenamefont {Bishop},\ and\ \citenamefont
  {Kubeka}}]{kakkat2024gravitational}%
  \BibitemOpen
  \bibfield  {author} {\bibinfo {author} {\bibfnamefont {V.}~\bibnamefont
  {Kakkat}}, \bibinfo {author} {\bibfnamefont {N.~T.}\ \bibnamefont {Bishop}},\
  and\ \bibinfo {author} {\bibfnamefont {A.~S.}\ \bibnamefont {Kubeka}},\
  }\bibfield  {title} {\bibinfo {title} {Gravitational wave heating},\
  }\href@noop {} {\bibfield  {journal} {\bibinfo  {journal} {Physical Review
  D}\ }\textbf {\bibinfo {volume} {109}},\ \bibinfo {pages} {024013} (\bibinfo
  {year} {2024})}\BibitemShut {NoStop}%
\bibitem [{\citenamefont {Milosavljevi{\'c}}\ and\ \citenamefont
  {Phinney}(2005)}]{milosavljevic2005afterglow}%
  \BibitemOpen
  \bibfield  {author} {\bibinfo {author} {\bibfnamefont {M.}~\bibnamefont
  {Milosavljevi{\'c}}}\ and\ \bibinfo {author} {\bibfnamefont {E.~S.}\
  \bibnamefont {Phinney}},\ }\bibfield  {title} {\bibinfo {title} {The
  afterglow of massive black hole coalescence},\ }\href@noop {} {\bibfield
  {journal} {\bibinfo  {journal} {The Astrophysical Journal}\ }\textbf
  {\bibinfo {volume} {622}},\ \bibinfo {pages} {L93} (\bibinfo {year}
  {2005})}\BibitemShut {NoStop}%
\bibitem [{\citenamefont {Tanaka}\ and\ \citenamefont
  {Menou}(2010)}]{tanaka2010time}%
  \BibitemOpen
  \bibfield  {author} {\bibinfo {author} {\bibfnamefont {T.}~\bibnamefont
  {Tanaka}}\ and\ \bibinfo {author} {\bibfnamefont {K.}~\bibnamefont {Menou}},\
  }\bibfield  {title} {\bibinfo {title} {Time-dependent models for the
  afterglows of massive black hole mergers},\ }\href@noop {} {\bibfield
  {journal} {\bibinfo  {journal} {The Astrophysical Journal}\ }\textbf
  {\bibinfo {volume} {714}},\ \bibinfo {pages} {404} (\bibinfo {year}
  {2010})}\BibitemShut {NoStop}%
\bibitem [{\citenamefont {Li}\ \emph {et~al.}(2012)\citenamefont {Li},
  \citenamefont {Kocsis},\ and\ \citenamefont {Loeb}}]{li2012gravitational}%
  \BibitemOpen
  \bibfield  {author} {\bibinfo {author} {\bibfnamefont {G.}~\bibnamefont
  {Li}}, \bibinfo {author} {\bibfnamefont {B.}~\bibnamefont {Kocsis}},\ and\
  \bibinfo {author} {\bibfnamefont {A.}~\bibnamefont {Loeb}},\ }\bibfield
  {title} {\bibinfo {title} {Gravitational wave heating of stars and accretion
  discs},\ }\href@noop {} {\bibfield  {journal} {\bibinfo  {journal} {Monthly
  Notices of the Royal Astronomical Society}\ }\textbf {\bibinfo {volume}
  {425}},\ \bibinfo {pages} {2407} (\bibinfo {year} {2012})}\BibitemShut
  {NoStop}%
\bibitem [{\citenamefont {Bishop}\ \emph {et~al.}(2024)\citenamefont {Bishop},
  \citenamefont {Kakkat}, \citenamefont {Kubeka}, \citenamefont {Naidoo},\ and\
  \citenamefont {van~der Walt}}]{Bishop2024essay}%
  \BibitemOpen
  \bibfield  {author} {\bibinfo {author} {\bibfnamefont {N.~T.}\ \bibnamefont
  {Bishop}}, \bibinfo {author} {\bibfnamefont {V.}~\bibnamefont {Kakkat}},
  \bibinfo {author} {\bibfnamefont {A.~S.}\ \bibnamefont {Kubeka}}, \bibinfo
  {author} {\bibfnamefont {M.}~\bibnamefont {Naidoo}},\ and\ \bibinfo {author}
  {\bibfnamefont {P.~J.}\ \bibnamefont {van~der Walt}},\ }\bibfield  {title}
  {\bibinfo {title} {The interaction of gravitational waves with matter},\
  }\href@noop {} {\bibfield  {journal} {\bibinfo  {journal} {arXiv preprint
  2405.07743}\ } (\bibinfo {year} {2024})}\BibitemShut {NoStop}%
\bibitem [{\citenamefont {Bishop}\ \emph {et~al.}(1999)\citenamefont {Bishop},
  \citenamefont {G{\'o}mez}, \citenamefont {Lehner}, \citenamefont {Maharaj},\
  and\ \citenamefont {Winicour}}]{bishop1999incorporation}%
  \BibitemOpen
  \bibfield  {author} {\bibinfo {author} {\bibfnamefont {N.~T.}\ \bibnamefont
  {Bishop}}, \bibinfo {author} {\bibfnamefont {R.}~\bibnamefont {G{\'o}mez}},
  \bibinfo {author} {\bibfnamefont {L.}~\bibnamefont {Lehner}}, \bibinfo
  {author} {\bibfnamefont {M.}~\bibnamefont {Maharaj}},\ and\ \bibinfo {author}
  {\bibfnamefont {J.}~\bibnamefont {Winicour}},\ }\bibfield  {title} {\bibinfo
  {title} {Incorporation of matter into characteristic numerical relativity},\
  }\href@noop {} {\bibfield  {journal} {\bibinfo  {journal} {Physical Review
  D}\ }\textbf {\bibinfo {volume} {60}},\ \bibinfo {pages} {024005} (\bibinfo
  {year} {1999})}\BibitemShut {NoStop}%
\bibitem [{\citenamefont {Bishop}\ \emph {et~al.}(1997)\citenamefont {Bishop},
  \citenamefont {G{\'o}mez}, \citenamefont {Lehner}, \citenamefont {Maharaj},\
  and\ \citenamefont {Winicour}}]{bishop1997high}%
  \BibitemOpen
  \bibfield  {author} {\bibinfo {author} {\bibfnamefont {N.~T.}\ \bibnamefont
  {Bishop}}, \bibinfo {author} {\bibfnamefont {R.}~\bibnamefont {G{\'o}mez}},
  \bibinfo {author} {\bibfnamefont {L.}~\bibnamefont {Lehner}}, \bibinfo
  {author} {\bibfnamefont {M.}~\bibnamefont {Maharaj}},\ and\ \bibinfo {author}
  {\bibfnamefont {J.}~\bibnamefont {Winicour}},\ }\bibfield  {title} {\bibinfo
  {title} {High-powered gravitational news},\ }\href@noop {} {\bibfield
  {journal} {\bibinfo  {journal} {Physical Review D}\ }\textbf {\bibinfo
  {volume} {56}},\ \bibinfo {pages} {6298} (\bibinfo {year}
  {1997})}\BibitemShut {NoStop}%
\bibitem [{\citenamefont {G{\'o}mez}(2001)}]{gomez2001gravitational}%
  \BibitemOpen
  \bibfield  {author} {\bibinfo {author} {\bibfnamefont {R.}~\bibnamefont
  {G{\'o}mez}},\ }\bibfield  {title} {\bibinfo {title} {Gravitational waveforms
  with controlled accuracy},\ }\href@noop {} {\bibfield  {journal} {\bibinfo
  {journal} {Physical Review D}\ }\textbf {\bibinfo {volume} {64}},\ \bibinfo
  {pages} {024007} (\bibinfo {year} {2001})}\BibitemShut {NoStop}%
\bibitem [{\citenamefont {Sachs}(1962)}]{sachs1962gravitational}%
  \BibitemOpen
  \bibfield  {author} {\bibinfo {author} {\bibfnamefont {R.~K.}\ \bibnamefont
  {Sachs}},\ }\bibfield  {title} {\bibinfo {title} {Gravitational waves in
  general relativity viii. waves in asymptotically flat space-time},\
  }\href@noop {} {\bibfield  {journal} {\bibinfo  {journal} {Proceedings of the
  Royal Society of London. Series A. Mathematical and Physical Sciences}\
  }\textbf {\bibinfo {volume} {270}},\ \bibinfo {pages} {103} (\bibinfo {year}
  {1962})}\BibitemShut {NoStop}%
\bibitem [{\citenamefont {Bondi}\ \emph {et~al.}(1962)\citenamefont {Bondi},
  \citenamefont {Van~der Burg},\ and\ \citenamefont
  {Metzner}}]{bondi1962gravitational}%
  \BibitemOpen
  \bibfield  {author} {\bibinfo {author} {\bibfnamefont {H.}~\bibnamefont
  {Bondi}}, \bibinfo {author} {\bibfnamefont {M.~G.~J.}\ \bibnamefont {Van~der
  Burg}},\ and\ \bibinfo {author} {\bibfnamefont {A.}~\bibnamefont {Metzner}},\
  }\bibfield  {title} {\bibinfo {title} {Gravitational waves in general
  relativity, vii. waves from axi-symmetric isolated system},\ }\href@noop {}
  {\bibfield  {journal} {\bibinfo  {journal} {Proceedings of the Royal Society
  of London. Series A. Mathematical and Physical Sciences}\ }\textbf {\bibinfo
  {volume} {269}},\ \bibinfo {pages} {21} (\bibinfo {year} {1962})}\BibitemShut
  {NoStop}%
\bibitem [{\citenamefont {G{\'o}mez}\ \emph {et~al.}(1997)\citenamefont
  {G{\'o}mez}, \citenamefont {Lehner}, \citenamefont {Papadopoulos},\ and\
  \citenamefont {Winicour}}]{gomez1997eth}%
  \BibitemOpen
  \bibfield  {author} {\bibinfo {author} {\bibfnamefont {R.}~\bibnamefont
  {G{\'o}mez}}, \bibinfo {author} {\bibfnamefont {L.}~\bibnamefont {Lehner}},
  \bibinfo {author} {\bibfnamefont {P.}~\bibnamefont {Papadopoulos}},\ and\
  \bibinfo {author} {\bibfnamefont {J.}~\bibnamefont {Winicour}},\ }\bibfield
  {title} {\bibinfo {title} {The eth formalism in numerical relativity},\
  }\href@noop {} {\bibfield  {journal} {\bibinfo  {journal} {Classical and
  Quantum Gravity}\ }\textbf {\bibinfo {volume} {14}},\ \bibinfo {pages} {977}
  (\bibinfo {year} {1997})}\BibitemShut {NoStop}%
\bibitem [{\citenamefont {Newman}\ and\ \citenamefont
  {Penrose}(1966)}]{newman1966note}%
  \BibitemOpen
  \bibfield  {author} {\bibinfo {author} {\bibfnamefont {E.~T.}\ \bibnamefont
  {Newman}}\ and\ \bibinfo {author} {\bibfnamefont {R.}~\bibnamefont
  {Penrose}},\ }\bibfield  {title} {\bibinfo {title} {Note on the
  bondi-metzner-sachs group},\ }\href@noop {} {\bibfield  {journal} {\bibinfo
  {journal} {Journal of Mathematical Physics}\ }\textbf {\bibinfo {volume}
  {7}},\ \bibinfo {pages} {863} (\bibinfo {year} {1966})}\BibitemShut {NoStop}%
\bibitem [{\citenamefont {Bishop}(2005)}]{bishop2005linearized}%
  \BibitemOpen
  \bibfield  {author} {\bibinfo {author} {\bibfnamefont {N.~T.}\ \bibnamefont
  {Bishop}},\ }\bibfield  {title} {\bibinfo {title} {Linearized solutions of
  the einstein equations within a bondi--sachs framework, and implications for
  boundary conditions in numerical simulations},\ }\href@noop {} {\bibfield
  {journal} {\bibinfo  {journal} {Classical and Quantum Gravity}\ }\textbf
  {\bibinfo {volume} {22}},\ \bibinfo {pages} {2393} (\bibinfo {year}
  {2005})}\BibitemShut {NoStop}%
\bibitem [{\citenamefont {Bishop}\ and\ \citenamefont
  {Rezzolla}(2016)}]{bishop2016extraction}%
  \BibitemOpen
  \bibfield  {author} {\bibinfo {author} {\bibfnamefont {N.~T.}\ \bibnamefont
  {Bishop}}\ and\ \bibinfo {author} {\bibfnamefont {L.}~\bibnamefont
  {Rezzolla}},\ }\bibfield  {title} {\bibinfo {title} {Extraction of
  gravitational waves in numerical relativity},\ }\href@noop {} {\bibfield
  {journal} {\bibinfo  {journal} {Living reviews in relativity}\ }\textbf
  {\bibinfo {volume} {19}},\ \bibinfo {pages} {1} (\bibinfo {year}
  {2016})}\BibitemShut {NoStop}%
\bibitem [{\citenamefont {Baumgarte}\ and\ \citenamefont
  {Shapiro}(2010)}]{baumgarte2010numerical}%
  \BibitemOpen
  \bibfield  {author} {\bibinfo {author} {\bibfnamefont {T.~W.}\ \bibnamefont
  {Baumgarte}}\ and\ \bibinfo {author} {\bibfnamefont {S.~L.}\ \bibnamefont
  {Shapiro}},\ }\href@noop {} {\emph {\bibinfo {title} {Numerical relativity:
  solving Einstein's equations on the computer}}}\ (\bibinfo  {publisher}
  {Cambridge University Press},\ \bibinfo {year} {2010})\BibitemShut {NoStop}%
\bibitem [{\citenamefont {Li}\ and\ \citenamefont
  {Paczy{\'n}ski}(1998)}]{li1998transient}%
  \BibitemOpen
  \bibfield  {author} {\bibinfo {author} {\bibfnamefont {L.-X.}\ \bibnamefont
  {Li}}\ and\ \bibinfo {author} {\bibfnamefont {B.}~\bibnamefont
  {Paczy{\'n}ski}},\ }\bibfield  {title} {\bibinfo {title} {Transient events
  from neutron star mergers},\ }\href@noop {} {\bibfield  {journal} {\bibinfo
  {journal} {The Astrophysical Journal}\ }\textbf {\bibinfo {volume} {507}},\
  \bibinfo {pages} {L59} (\bibinfo {year} {1998})}\BibitemShut {NoStop}%
\bibitem [{\citenamefont {Metzger}\ \emph {et~al.}(2010)\citenamefont
  {Metzger}, \citenamefont {Mart{\'\i}nez-Pinedo}, \citenamefont {Darbha},
  \citenamefont {Quataert}, \citenamefont {Arcones}, \citenamefont {Kasen},
  \citenamefont {Thomas}, \citenamefont {Nugent}, \citenamefont {Panov},\ and\
  \citenamefont {Zinner}}]{metzger2010electromagnetic}%
  \BibitemOpen
  \bibfield  {author} {\bibinfo {author} {\bibfnamefont {B.}~\bibnamefont
  {Metzger}}, \bibinfo {author} {\bibfnamefont {G.}~\bibnamefont
  {Mart{\'\i}nez-Pinedo}}, \bibinfo {author} {\bibfnamefont {S.}~\bibnamefont
  {Darbha}}, \bibinfo {author} {\bibfnamefont {E.}~\bibnamefont {Quataert}},
  \bibinfo {author} {\bibfnamefont {A.}~\bibnamefont {Arcones}}, \bibinfo
  {author} {\bibfnamefont {D.}~\bibnamefont {Kasen}}, \bibinfo {author}
  {\bibfnamefont {R.}~\bibnamefont {Thomas}}, \bibinfo {author} {\bibfnamefont
  {P.}~\bibnamefont {Nugent}}, \bibinfo {author} {\bibfnamefont
  {I.}~\bibnamefont {Panov}},\ and\ \bibinfo {author} {\bibfnamefont
  {N.}~\bibnamefont {Zinner}},\ }\bibfield  {title} {\bibinfo {title}
  {Electromagnetic counterparts of compact object mergers powered by the
  radioactive decay of r-process nuclei},\ }\href@noop {} {\bibfield  {journal}
  {\bibinfo  {journal} {Monthly Notices of the Royal Astronomical Society}\
  }\textbf {\bibinfo {volume} {406}},\ \bibinfo {pages} {2650} (\bibinfo {year}
  {2010})}\BibitemShut {NoStop}%
\bibitem [{\citenamefont {Roberts}\ \emph {et~al.}(2011)\citenamefont
  {Roberts}, \citenamefont {Kasen}, \citenamefont {Lee},\ and\ \citenamefont
  {Ramirez-Ruiz}}]{roberts2011electromagnetic}%
  \BibitemOpen
  \bibfield  {author} {\bibinfo {author} {\bibfnamefont {L.~F.}\ \bibnamefont
  {Roberts}}, \bibinfo {author} {\bibfnamefont {D.}~\bibnamefont {Kasen}},
  \bibinfo {author} {\bibfnamefont {W.~H.}\ \bibnamefont {Lee}},\ and\ \bibinfo
  {author} {\bibfnamefont {E.}~\bibnamefont {Ramirez-Ruiz}},\ }\bibfield
  {title} {\bibinfo {title} {Electromagnetic transients powered by nuclear
  decay in the tidal tails of coalescing compact binaries},\ }\href@noop {}
  {\bibfield  {journal} {\bibinfo  {journal} {The Astrophysical Journal
  Letters}\ }\textbf {\bibinfo {volume} {736}},\ \bibinfo {pages} {L21}
  (\bibinfo {year} {2011})}\BibitemShut {NoStop}%
\bibitem [{\citenamefont {Barnes}\ and\ \citenamefont
  {Kasen}(2013)}]{barnes2013effect}%
  \BibitemOpen
  \bibfield  {author} {\bibinfo {author} {\bibfnamefont {J.}~\bibnamefont
  {Barnes}}\ and\ \bibinfo {author} {\bibfnamefont {D.}~\bibnamefont {Kasen}},\
  }\bibfield  {title} {\bibinfo {title} {Effect of a high opacity on the light
  curves of radioactively powered transients from compact object mergers},\
  }\href@noop {} {\bibfield  {journal} {\bibinfo  {journal} {The Astrophysical
  Journal}\ }\textbf {\bibinfo {volume} {775}},\ \bibinfo {pages} {18}
  (\bibinfo {year} {2013})}\BibitemShut {NoStop}%
\bibitem [{\citenamefont {Tanaka}\ and\ \citenamefont
  {Hotokezaka}(2013)}]{tanaka2013radiative}%
  \BibitemOpen
  \bibfield  {author} {\bibinfo {author} {\bibfnamefont {M.}~\bibnamefont
  {Tanaka}}\ and\ \bibinfo {author} {\bibfnamefont {K.}~\bibnamefont
  {Hotokezaka}},\ }\bibfield  {title} {\bibinfo {title} {Radiative transfer
  simulations of neutron star merger ejecta},\ }\href@noop {} {\bibfield
  {journal} {\bibinfo  {journal} {The Astrophysical Journal}\ }\textbf
  {\bibinfo {volume} {775}},\ \bibinfo {pages} {113} (\bibinfo {year}
  {2013})}\BibitemShut {NoStop}%
\bibitem [{\citenamefont {Abbott}\ \emph {et~al.}(2017)\citenamefont {Abbott}
  \emph {et~al.}}]{abbott2017multi}%
  \BibitemOpen
  \bibfield  {author} {\bibinfo {author} {\bibfnamefont {B.~P.}\ \bibnamefont
  {Abbott}} \emph {et~al.} (\bibinfo {collaboration} {LIGO Scientific, Virgo,
  Fermi GBM, INTEGRAL, IceCube, AstroSat Cadmium Zinc Telluride Imager Team,
  IPN, Insight-Hxmt, ANTARES, Swift, AGILE Team, 1M2H Team, Dark Energy Camera
  GW-EM, DES, DLT40, GRAWITA, Fermi-LAT, ATCA, ASKAP, Las Cumbres Observatory
  Group, OzGrav, DWF (Deeper Wider Faster Program), AST3, CAASTRO, VINROUGE,
  MASTER, J-GEM, GROWTH, JAGWAR, CaltechNRAO, TTU-NRAO, NuSTAR, Pan-STARRS,
  MAXI Team, TZAC Consortium, KU, Nordic Optical Telescope, ePESSTO, GROND,
  Texas Tech University, SALT Group, TOROS, BOOTES, MWA, CALET, IKI-GW
  Follow-up, H.E.S.S., LOFAR, LWA, HAWC, Pierre Auger, ALMA, Euro VLBI Team, Pi
  of Sky, Chandra Team at McGill University, DFN, ATLAS Telescopes, High Time
  Resolution Universe Survey, RIMAS, RATIR, SKA South Africa/MeerKAT}),\
  }\bibfield  {title} {\bibinfo {title} {{Multi-messenger Observations of a
  Binary Neutron Star Merger}},\ }\href
  {https://doi.org/10.3847/2041-8213/aa91c9} {\bibfield  {journal} {\bibinfo
  {journal} {Astrophys. J. Lett.}\ }\textbf {\bibinfo {volume} {848}},\
  \bibinfo {pages} {L12} (\bibinfo {year} {2017})},\ \Eprint
  {https://arxiv.org/abs/1710.05833} {arXiv:1710.05833 [astro-ph.HE]}
  \BibitemShut {NoStop}%
\bibitem [{\citenamefont {Tanaka}\ \emph {et~al.}(2017)\citenamefont {Tanaka},
  \citenamefont {Utsumi}, \citenamefont {Mazzali}, \citenamefont {Tominaga},
  \citenamefont {Yoshida}, \citenamefont {Sekiguchi}, \citenamefont {Morokuma},
  \citenamefont {Motohara}, \citenamefont {Ohta}, \citenamefont {Kawabata}
  \emph {et~al.}}]{tanaka2017kilonova}%
  \BibitemOpen
  \bibfield  {author} {\bibinfo {author} {\bibfnamefont {M.}~\bibnamefont
  {Tanaka}}, \bibinfo {author} {\bibfnamefont {Y.}~\bibnamefont {Utsumi}},
  \bibinfo {author} {\bibfnamefont {P.~A.}\ \bibnamefont {Mazzali}}, \bibinfo
  {author} {\bibfnamefont {N.}~\bibnamefont {Tominaga}}, \bibinfo {author}
  {\bibfnamefont {M.}~\bibnamefont {Yoshida}}, \bibinfo {author} {\bibfnamefont
  {Y.}~\bibnamefont {Sekiguchi}}, \bibinfo {author} {\bibfnamefont
  {T.}~\bibnamefont {Morokuma}}, \bibinfo {author} {\bibfnamefont
  {K.}~\bibnamefont {Motohara}}, \bibinfo {author} {\bibfnamefont
  {K.}~\bibnamefont {Ohta}}, \bibinfo {author} {\bibfnamefont {K.~S.}\
  \bibnamefont {Kawabata}}, \emph {et~al.},\ }\bibfield  {title} {\bibinfo
  {title} {Kilonova from post-merger ejecta as an optical and near-infrared
  counterpart of gw170817},\ }\href@noop {} {\bibfield  {journal} {\bibinfo
  {journal} {Publications of the Astronomical Society of Japan}\ }\textbf
  {\bibinfo {volume} {69}},\ \bibinfo {pages} {102} (\bibinfo {year}
  {2017})}\BibitemShut {NoStop}%
\bibitem [{\citenamefont {Arcavi}\ \emph {et~al.}(2017)\citenamefont {Arcavi},
  \citenamefont {Hosseinzadeh}, \citenamefont {Howell}, \citenamefont
  {McCully}, \citenamefont {Poznanski}, \citenamefont {Kasen}, \citenamefont
  {Barnes}, \citenamefont {Zaltzman}, \citenamefont {Vasylyev}, \citenamefont
  {Maoz} \emph {et~al.}}]{arcavi2017optical}%
  \BibitemOpen
  \bibfield  {author} {\bibinfo {author} {\bibfnamefont {I.}~\bibnamefont
  {Arcavi}}, \bibinfo {author} {\bibfnamefont {G.}~\bibnamefont
  {Hosseinzadeh}}, \bibinfo {author} {\bibfnamefont {D.~A.}\ \bibnamefont
  {Howell}}, \bibinfo {author} {\bibfnamefont {C.}~\bibnamefont {McCully}},
  \bibinfo {author} {\bibfnamefont {D.}~\bibnamefont {Poznanski}}, \bibinfo
  {author} {\bibfnamefont {D.}~\bibnamefont {Kasen}}, \bibinfo {author}
  {\bibfnamefont {J.}~\bibnamefont {Barnes}}, \bibinfo {author} {\bibfnamefont
  {M.}~\bibnamefont {Zaltzman}}, \bibinfo {author} {\bibfnamefont
  {S.}~\bibnamefont {Vasylyev}}, \bibinfo {author} {\bibfnamefont
  {D.}~\bibnamefont {Maoz}}, \emph {et~al.},\ }\bibfield  {title} {\bibinfo
  {title} {Optical emission from a kilonova following a
  gravitational-wave-detected neutron-star merger},\ }\href@noop {} {\bibfield
  {journal} {\bibinfo  {journal} {Nature}\ }\textbf {\bibinfo {volume} {551}},\
  \bibinfo {pages} {64} (\bibinfo {year} {2017})}\BibitemShut {NoStop}%
\bibitem [{\citenamefont {Coulter}\ \emph {et~al.}(2017)\citenamefont
  {Coulter}, \citenamefont {Foley}, \citenamefont {Kilpatrick}, \citenamefont
  {Drout}, \citenamefont {Piro}, \citenamefont {Shappee}, \citenamefont
  {Siebert}, \citenamefont {Simon}, \citenamefont {Ulloa}, \citenamefont
  {Kasen} \emph {et~al.}}]{coulter2017swope}%
  \BibitemOpen
  \bibfield  {author} {\bibinfo {author} {\bibfnamefont {D.}~\bibnamefont
  {Coulter}}, \bibinfo {author} {\bibfnamefont {R.}~\bibnamefont {Foley}},
  \bibinfo {author} {\bibfnamefont {C.}~\bibnamefont {Kilpatrick}}, \bibinfo
  {author} {\bibfnamefont {M.}~\bibnamefont {Drout}}, \bibinfo {author}
  {\bibfnamefont {A.}~\bibnamefont {Piro}}, \bibinfo {author} {\bibfnamefont
  {B.}~\bibnamefont {Shappee}}, \bibinfo {author} {\bibfnamefont
  {M.}~\bibnamefont {Siebert}}, \bibinfo {author} {\bibfnamefont
  {J.}~\bibnamefont {Simon}}, \bibinfo {author} {\bibfnamefont
  {N.}~\bibnamefont {Ulloa}}, \bibinfo {author} {\bibfnamefont
  {D.}~\bibnamefont {Kasen}}, \emph {et~al.},\ }\bibfield  {title} {\bibinfo
  {title} {Swope supernova survey 2017a (sss17a), the optical counterpart to a
  gravitational wave source},\ }\href@noop {} {\bibfield  {journal} {\bibinfo
  {journal} {Science}\ }\textbf {\bibinfo {volume} {358}},\ \bibinfo {pages}
  {1556} (\bibinfo {year} {2017})}\BibitemShut {NoStop}%
\bibitem [{\citenamefont {Cowperthwaite}\ \emph {et~al.}(2017)\citenamefont
  {Cowperthwaite}, \citenamefont {Berger}, \citenamefont {Villar},
  \citenamefont {Metzger}, \citenamefont {Nicholl}, \citenamefont {Chornock},
  \citenamefont {Blanchard}, \citenamefont {Fong}, \citenamefont {Margutti},
  \citenamefont {Soares-Santos} \emph
  {et~al.}}]{cowperthwaite2017electromagnetic}%
  \BibitemOpen
  \bibfield  {author} {\bibinfo {author} {\bibfnamefont {P.~S.}\ \bibnamefont
  {Cowperthwaite}}, \bibinfo {author} {\bibfnamefont {E.}~\bibnamefont
  {Berger}}, \bibinfo {author} {\bibfnamefont {V.}~\bibnamefont {Villar}},
  \bibinfo {author} {\bibfnamefont {B.}~\bibnamefont {Metzger}}, \bibinfo
  {author} {\bibfnamefont {M.}~\bibnamefont {Nicholl}}, \bibinfo {author}
  {\bibfnamefont {R.}~\bibnamefont {Chornock}}, \bibinfo {author}
  {\bibfnamefont {P.}~\bibnamefont {Blanchard}}, \bibinfo {author}
  {\bibfnamefont {W.~f.}\ \bibnamefont {Fong}}, \bibinfo {author}
  {\bibfnamefont {R.}~\bibnamefont {Margutti}}, \bibinfo {author}
  {\bibfnamefont {M.}~\bibnamefont {Soares-Santos}}, \emph {et~al.},\
  }\bibfield  {title} {\bibinfo {title} {The electromagnetic counterpart of the
  binary neutron star merger ligo/virgo gw170817. ii. uv, optical, and
  near-infrared light curves and comparison to kilonova models},\ }\href@noop
  {} {\bibfield  {journal} {\bibinfo  {journal} {The Astrophysical Journal
  Letters}\ }\textbf {\bibinfo {volume} {848}},\ \bibinfo {pages} {L17}
  (\bibinfo {year} {2017})}\BibitemShut {NoStop}%
\bibitem [{\citenamefont {Drout}\ \emph {et~al.}(2017)\citenamefont {Drout},
  \citenamefont {Piro}, \citenamefont {Shappee}, \citenamefont {Kilpatrick},
  \citenamefont {Simon}, \citenamefont {Contreras}, \citenamefont {Coulter},
  \citenamefont {Foley}, \citenamefont {Siebert}, \citenamefont {Morrell} \emph
  {et~al.}}]{drout2017light}%
  \BibitemOpen
  \bibfield  {author} {\bibinfo {author} {\bibfnamefont {M.~R.}\ \bibnamefont
  {Drout}}, \bibinfo {author} {\bibfnamefont {A.}~\bibnamefont {Piro}},
  \bibinfo {author} {\bibfnamefont {B.}~\bibnamefont {Shappee}}, \bibinfo
  {author} {\bibfnamefont {C.}~\bibnamefont {Kilpatrick}}, \bibinfo {author}
  {\bibfnamefont {J.}~\bibnamefont {Simon}}, \bibinfo {author} {\bibfnamefont
  {C.}~\bibnamefont {Contreras}}, \bibinfo {author} {\bibfnamefont
  {D.}~\bibnamefont {Coulter}}, \bibinfo {author} {\bibfnamefont
  {R.}~\bibnamefont {Foley}}, \bibinfo {author} {\bibfnamefont
  {M.}~\bibnamefont {Siebert}}, \bibinfo {author} {\bibfnamefont
  {N.}~\bibnamefont {Morrell}}, \emph {et~al.},\ }\bibfield  {title} {\bibinfo
  {title} {Light curves of the neutron star merger gw170817/sss17a:
  Implications for r-process nucleosynthesis},\ }\href@noop {} {\bibfield
  {journal} {\bibinfo  {journal} {Science}\ }\textbf {\bibinfo {volume}
  {358}},\ \bibinfo {pages} {1570} (\bibinfo {year} {2017})}\BibitemShut
  {NoStop}%
\bibitem [{\citenamefont {Evans}\ \emph {et~al.}(2017)\citenamefont {Evans},
  \citenamefont {Cenko}, \citenamefont {Kennea}, \citenamefont {Emery},
  \citenamefont {Kuin}, \citenamefont {Korobkin}, \citenamefont {Wollaeger},
  \citenamefont {Fryer}, \citenamefont {Madsen}, \citenamefont {Harrison} \emph
  {et~al.}}]{evans2017swift}%
  \BibitemOpen
  \bibfield  {author} {\bibinfo {author} {\bibfnamefont {P.}~\bibnamefont
  {Evans}}, \bibinfo {author} {\bibfnamefont {S.}~\bibnamefont {Cenko}},
  \bibinfo {author} {\bibfnamefont {J.}~\bibnamefont {Kennea}}, \bibinfo
  {author} {\bibfnamefont {S.}~\bibnamefont {Emery}}, \bibinfo {author}
  {\bibfnamefont {N.}~\bibnamefont {Kuin}}, \bibinfo {author} {\bibfnamefont
  {O.}~\bibnamefont {Korobkin}}, \bibinfo {author} {\bibfnamefont
  {R.}~\bibnamefont {Wollaeger}}, \bibinfo {author} {\bibfnamefont
  {C.}~\bibnamefont {Fryer}}, \bibinfo {author} {\bibfnamefont
  {K.}~\bibnamefont {Madsen}}, \bibinfo {author} {\bibfnamefont
  {F.}~\bibnamefont {Harrison}}, \emph {et~al.},\ }\bibfield  {title} {\bibinfo
  {title} {Swift and nustar observations of gw170817: detection of a blue
  kilonova},\ }\href@noop {} {\bibfield  {journal} {\bibinfo  {journal}
  {Science}\ }\textbf {\bibinfo {volume} {358}},\ \bibinfo {pages} {1565}
  (\bibinfo {year} {2017})}\BibitemShut {NoStop}%
\bibitem [{\citenamefont {Kasliwal}\ \emph {et~al.}(2017)\citenamefont
  {Kasliwal}, \citenamefont {Nakar}, \citenamefont {Singer}, \citenamefont
  {Kaplan}, \citenamefont {Cook}, \citenamefont {Van~Sistine}, \citenamefont
  {Lau}, \citenamefont {Fremling}, \citenamefont {Gottlieb}, \citenamefont
  {Jencson} \emph {et~al.}}]{kasliwal2017illuminating}%
  \BibitemOpen
  \bibfield  {author} {\bibinfo {author} {\bibfnamefont {M.}~\bibnamefont
  {Kasliwal}}, \bibinfo {author} {\bibfnamefont {E.}~\bibnamefont {Nakar}},
  \bibinfo {author} {\bibfnamefont {L.}~\bibnamefont {Singer}}, \bibinfo
  {author} {\bibfnamefont {D.}~\bibnamefont {Kaplan}}, \bibinfo {author}
  {\bibfnamefont {D.}~\bibnamefont {Cook}}, \bibinfo {author} {\bibfnamefont
  {A.}~\bibnamefont {Van~Sistine}}, \bibinfo {author} {\bibfnamefont
  {R.}~\bibnamefont {Lau}}, \bibinfo {author} {\bibfnamefont {C.}~\bibnamefont
  {Fremling}}, \bibinfo {author} {\bibfnamefont {O.}~\bibnamefont {Gottlieb}},
  \bibinfo {author} {\bibfnamefont {J.}~\bibnamefont {Jencson}}, \emph
  {et~al.},\ }\bibfield  {title} {\bibinfo {title} {Illuminating gravitational
  waves: A concordant picture of photons from a neutron star merger},\
  }\href@noop {} {\bibfield  {journal} {\bibinfo  {journal} {Science}\ }\textbf
  {\bibinfo {volume} {358}},\ \bibinfo {pages} {1559} (\bibinfo {year}
  {2017})}\BibitemShut {NoStop}%
\bibitem [{\citenamefont {Smartt}\ \emph {et~al.}(2017)\citenamefont {Smartt},
  \citenamefont {Chen}, \citenamefont {Jerkstrand}, \citenamefont {Coughlin},
  \citenamefont {Kankare}, \citenamefont {Sim}, \citenamefont {Fraser},
  \citenamefont {Inserra}, \citenamefont {Maguire}, \citenamefont {Chambers}
  \emph {et~al.}}]{smartt2017kilonova}%
  \BibitemOpen
  \bibfield  {author} {\bibinfo {author} {\bibfnamefont {S.}~\bibnamefont
  {Smartt}}, \bibinfo {author} {\bibfnamefont {T.-W.}\ \bibnamefont {Chen}},
  \bibinfo {author} {\bibfnamefont {A.}~\bibnamefont {Jerkstrand}}, \bibinfo
  {author} {\bibfnamefont {M.}~\bibnamefont {Coughlin}}, \bibinfo {author}
  {\bibfnamefont {E.}~\bibnamefont {Kankare}}, \bibinfo {author} {\bibfnamefont
  {S.}~\bibnamefont {Sim}}, \bibinfo {author} {\bibfnamefont {M.}~\bibnamefont
  {Fraser}}, \bibinfo {author} {\bibfnamefont {C.}~\bibnamefont {Inserra}},
  \bibinfo {author} {\bibfnamefont {K.}~\bibnamefont {Maguire}}, \bibinfo
  {author} {\bibfnamefont {K.}~\bibnamefont {Chambers}}, \emph {et~al.},\
  }\bibfield  {title} {\bibinfo {title} {A kilonova as the electromagnetic
  counterpart to a gravitational-wave source},\ }\href@noop {} {\bibfield
  {journal} {\bibinfo  {journal} {Nature}\ }\textbf {\bibinfo {volume} {551}},\
  \bibinfo {pages} {75} (\bibinfo {year} {2017})}\BibitemShut {NoStop}%
\bibitem [{\citenamefont {Tanvir}\ \emph {et~al.}(2017)\citenamefont {Tanvir},
  \citenamefont {Levan}, \citenamefont {Gonz{\'a}lez-Fern{\'a}ndez},
  \citenamefont {Korobkin}, \citenamefont {Mandel}, \citenamefont {Rosswog},
  \citenamefont {Hjorth}, \citenamefont {D’Avanzo}, \citenamefont {Fruchter},
  \citenamefont {Fryer} \emph {et~al.}}]{tanvir2017emergence}%
  \BibitemOpen
  \bibfield  {author} {\bibinfo {author} {\bibfnamefont {N.~R.}\ \bibnamefont
  {Tanvir}}, \bibinfo {author} {\bibfnamefont {A.~J.}\ \bibnamefont {Levan}},
  \bibinfo {author} {\bibfnamefont {C.}~\bibnamefont
  {Gonz{\'a}lez-Fern{\'a}ndez}}, \bibinfo {author} {\bibfnamefont
  {O.}~\bibnamefont {Korobkin}}, \bibinfo {author} {\bibfnamefont
  {I.}~\bibnamefont {Mandel}}, \bibinfo {author} {\bibfnamefont
  {S.}~\bibnamefont {Rosswog}}, \bibinfo {author} {\bibfnamefont
  {J.}~\bibnamefont {Hjorth}}, \bibinfo {author} {\bibfnamefont
  {P.}~\bibnamefont {D’Avanzo}}, \bibinfo {author} {\bibfnamefont
  {A.}~\bibnamefont {Fruchter}}, \bibinfo {author} {\bibfnamefont
  {C.}~\bibnamefont {Fryer}}, \emph {et~al.},\ }\bibfield  {title} {\bibinfo
  {title} {The emergence of a lanthanide-rich kilonova following the merger of
  two neutron stars},\ }\href@noop {} {\bibfield  {journal} {\bibinfo
  {journal} {The Astrophysical Journal Letters}\ }\textbf {\bibinfo {volume}
  {848}},\ \bibinfo {pages} {L27} (\bibinfo {year} {2017})}\BibitemShut
  {NoStop}%
\bibitem [{\citenamefont {Rosswog}\ \emph {et~al.}(1998)\citenamefont
  {Rosswog}, \citenamefont {Liebend{\"o}rfer}, \citenamefont {Thielemann},
  \citenamefont {Davies}, \citenamefont {Benz},\ and\ \citenamefont
  {Piran}}]{rosswog1998mass}%
  \BibitemOpen
  \bibfield  {author} {\bibinfo {author} {\bibfnamefont {S.}~\bibnamefont
  {Rosswog}}, \bibinfo {author} {\bibfnamefont {M.}~\bibnamefont
  {Liebend{\"o}rfer}}, \bibinfo {author} {\bibfnamefont {F.-K.}\ \bibnamefont
  {Thielemann}}, \bibinfo {author} {\bibfnamefont {M.}~\bibnamefont {Davies}},
  \bibinfo {author} {\bibfnamefont {W.}~\bibnamefont {Benz}},\ and\ \bibinfo
  {author} {\bibfnamefont {T.}~\bibnamefont {Piran}},\ }\bibfield  {title}
  {\bibinfo {title} {Mass ejection in neutron star mergers},\ }\href@noop {}
  {\bibfield  {journal} {\bibinfo  {journal} {Arxiv preprint astro-ph/9811367}\
  } (\bibinfo {year} {1998})}\BibitemShut {NoStop}%
\bibitem [{\citenamefont {Ruffert}\ \emph {et~al.}(1996)\citenamefont
  {Ruffert}, \citenamefont {Janka}, \citenamefont {Takahashi},\ and\
  \citenamefont {Schaefer}}]{ruffert1996coalescing}%
  \BibitemOpen
  \bibfield  {author} {\bibinfo {author} {\bibfnamefont {M.}~\bibnamefont
  {Ruffert}}, \bibinfo {author} {\bibfnamefont {H.-T.}\ \bibnamefont {Janka}},
  \bibinfo {author} {\bibfnamefont {K.}~\bibnamefont {Takahashi}},\ and\
  \bibinfo {author} {\bibfnamefont {G.}~\bibnamefont {Schaefer}},\ }\bibfield
  {title} {\bibinfo {title} {Coalescing neutron stars--a step towards physical
  models. ii. neutrino emission, neutron tori, and gamma-ray bursts},\
  }\href@noop {} {\bibfield  {journal} {\bibinfo  {journal} {arXiv preprint
  astro-ph/9606181}\ } (\bibinfo {year} {1996})}\BibitemShut {NoStop}%
\bibitem [{\citenamefont {Freiburghaus}\ \emph {et~al.}(1999)\citenamefont
  {Freiburghaus}, \citenamefont {Rosswog},\ and\ \citenamefont
  {Thielemann}}]{freiburghaus1999r}%
  \BibitemOpen
  \bibfield  {author} {\bibinfo {author} {\bibfnamefont {C.}~\bibnamefont
  {Freiburghaus}}, \bibinfo {author} {\bibfnamefont {S.}~\bibnamefont
  {Rosswog}},\ and\ \bibinfo {author} {\bibfnamefont {F.-K.}\ \bibnamefont
  {Thielemann}},\ }\bibfield  {title} {\bibinfo {title} {R-process in neutron
  star mergers},\ }\href@noop {} {\bibfield  {journal} {\bibinfo  {journal}
  {The Astrophysical Journal}\ }\textbf {\bibinfo {volume} {525}},\ \bibinfo
  {pages} {L121} (\bibinfo {year} {1999})}\BibitemShut {NoStop}%
\bibitem [{\citenamefont {Fern{\'a}ndez}\ and\ \citenamefont
  {Metzger}(2013)}]{fernandez2013delayed}%
  \BibitemOpen
  \bibfield  {author} {\bibinfo {author} {\bibfnamefont {R.}~\bibnamefont
  {Fern{\'a}ndez}}\ and\ \bibinfo {author} {\bibfnamefont {B.~D.}\ \bibnamefont
  {Metzger}},\ }\bibfield  {title} {\bibinfo {title} {Delayed outflows from
  black hole accretion tori following neutron star binary coalescence},\
  }\href@noop {} {\bibfield  {journal} {\bibinfo  {journal} {Monthly Notices of
  the Royal Astronomical Society}\ }\textbf {\bibinfo {volume} {435}},\
  \bibinfo {pages} {502} (\bibinfo {year} {2013})}\BibitemShut {NoStop}%
\bibitem [{\citenamefont {Metzger}\ and\ \citenamefont
  {Fern{\'a}ndez}(2014)}]{metzger2014red}%
  \BibitemOpen
  \bibfield  {author} {\bibinfo {author} {\bibfnamefont {B.~D.}\ \bibnamefont
  {Metzger}}\ and\ \bibinfo {author} {\bibfnamefont {R.}~\bibnamefont
  {Fern{\'a}ndez}},\ }\bibfield  {title} {\bibinfo {title} {Red or blue? a
  potential kilonova imprint of the delay until black hole formation following
  a neutron star merger},\ }\href@noop {} {\bibfield  {journal} {\bibinfo
  {journal} {Monthly Notices of the Royal Astronomical Society}\ }\textbf
  {\bibinfo {volume} {441}},\ \bibinfo {pages} {3444} (\bibinfo {year}
  {2014})}\BibitemShut {NoStop}%
\bibitem [{\citenamefont {Perego}\ \emph {et~al.}(2014)\citenamefont {Perego},
  \citenamefont {Rosswog}, \citenamefont {Cabez{\'o}n}, \citenamefont
  {Korobkin}, \citenamefont {K{\"a}ppeli}, \citenamefont {Arcones},\ and\
  \citenamefont {Liebend{\"o}rfer}}]{perego2014neutrino}%
  \BibitemOpen
  \bibfield  {author} {\bibinfo {author} {\bibfnamefont {A.}~\bibnamefont
  {Perego}}, \bibinfo {author} {\bibfnamefont {S.}~\bibnamefont {Rosswog}},
  \bibinfo {author} {\bibfnamefont {R.~M.}\ \bibnamefont {Cabez{\'o}n}},
  \bibinfo {author} {\bibfnamefont {O.}~\bibnamefont {Korobkin}}, \bibinfo
  {author} {\bibfnamefont {R.}~\bibnamefont {K{\"a}ppeli}}, \bibinfo {author}
  {\bibfnamefont {A.}~\bibnamefont {Arcones}},\ and\ \bibinfo {author}
  {\bibfnamefont {M.}~\bibnamefont {Liebend{\"o}rfer}},\ }\bibfield  {title}
  {\bibinfo {title} {Neutrino-driven winds from neutron star merger remnants},\
  }\href@noop {} {\bibfield  {journal} {\bibinfo  {journal} {Monthly Notices of
  the Royal Astronomical Society}\ }\textbf {\bibinfo {volume} {443}},\
  \bibinfo {pages} {3134} (\bibinfo {year} {2014})}\BibitemShut {NoStop}%
\bibitem [{\citenamefont {Fern{\'a}ndez}\ \emph {et~al.}(2015)\citenamefont
  {Fern{\'a}ndez}, \citenamefont {Kasen}, \citenamefont {Metzger},\ and\
  \citenamefont {Quataert}}]{fernandez2015outflows}%
  \BibitemOpen
  \bibfield  {author} {\bibinfo {author} {\bibfnamefont {R.}~\bibnamefont
  {Fern{\'a}ndez}}, \bibinfo {author} {\bibfnamefont {D.}~\bibnamefont
  {Kasen}}, \bibinfo {author} {\bibfnamefont {B.~D.}\ \bibnamefont {Metzger}},\
  and\ \bibinfo {author} {\bibfnamefont {E.}~\bibnamefont {Quataert}},\
  }\bibfield  {title} {\bibinfo {title} {Outflows from accretion discs formed
  in neutron star mergers: effect of black hole spin},\ }\href@noop {}
  {\bibfield  {journal} {\bibinfo  {journal} {Monthly Notices of the Royal
  Astronomical Society}\ }\textbf {\bibinfo {volume} {446}},\ \bibinfo {pages}
  {750} (\bibinfo {year} {2015})}\BibitemShut {NoStop}%
\bibitem [{\citenamefont {Just}\ \emph {et~al.}(2015)\citenamefont {Just},
  \citenamefont {Bauswein}, \citenamefont {Pulpillo}, \citenamefont {Goriely},\
  and\ \citenamefont {Janka}}]{just2015comprehensive}%
  \BibitemOpen
  \bibfield  {author} {\bibinfo {author} {\bibfnamefont {O.}~\bibnamefont
  {Just}}, \bibinfo {author} {\bibfnamefont {A.}~\bibnamefont {Bauswein}},
  \bibinfo {author} {\bibfnamefont {R.~A.}\ \bibnamefont {Pulpillo}}, \bibinfo
  {author} {\bibfnamefont {S.}~\bibnamefont {Goriely}},\ and\ \bibinfo {author}
  {\bibfnamefont {H.-T.}\ \bibnamefont {Janka}},\ }\bibfield  {title} {\bibinfo
  {title} {Comprehensive nucleosynthesis analysis for ejecta of compact binary
  mergers},\ }\href@noop {} {\bibfield  {journal} {\bibinfo  {journal} {Monthly
  Notices of the Royal Astronomical Society}\ }\textbf {\bibinfo {volume}
  {448}},\ \bibinfo {pages} {541} (\bibinfo {year} {2015})}\BibitemShut
  {NoStop}%
\bibitem [{\citenamefont {Siegel}\ and\ \citenamefont
  {Metzger}(2017)}]{siegel2017three}%
  \BibitemOpen
  \bibfield  {author} {\bibinfo {author} {\bibfnamefont {D.~M.}\ \bibnamefont
  {Siegel}}\ and\ \bibinfo {author} {\bibfnamefont {B.~D.}\ \bibnamefont
  {Metzger}},\ }\bibfield  {title} {\bibinfo {title} {Three-dimensional
  general-relativistic magnetohydrodynamic simulations of remnant accretion
  disks from neutron star mergers: outflows and r-process nucleosynthesis},\
  }\href@noop {} {\bibfield  {journal} {\bibinfo  {journal} {Physical Review
  Letters}\ }\textbf {\bibinfo {volume} {119}},\ \bibinfo {pages} {231102}
  (\bibinfo {year} {2017})}\BibitemShut {NoStop}%
\bibitem [{\citenamefont {Fujibayashi}\ \emph {et~al.}(2018)\citenamefont
  {Fujibayashi}, \citenamefont {Kiuchi}, \citenamefont {Nishimura},
  \citenamefont {Sekiguchi},\ and\ \citenamefont
  {Shibata}}]{fujibayashi2018mass}%
  \BibitemOpen
  \bibfield  {author} {\bibinfo {author} {\bibfnamefont {S.}~\bibnamefont
  {Fujibayashi}}, \bibinfo {author} {\bibfnamefont {K.}~\bibnamefont {Kiuchi}},
  \bibinfo {author} {\bibfnamefont {N.}~\bibnamefont {Nishimura}}, \bibinfo
  {author} {\bibfnamefont {Y.}~\bibnamefont {Sekiguchi}},\ and\ \bibinfo
  {author} {\bibfnamefont {M.}~\bibnamefont {Shibata}},\ }\bibfield  {title}
  {\bibinfo {title} {Mass ejection from the remnant of a binary neutron star
  merger: viscous-radiation hydrodynamics study},\ }\href@noop {} {\bibfield
  {journal} {\bibinfo  {journal} {The Astrophysical Journal}\ }\textbf
  {\bibinfo {volume} {860}},\ \bibinfo {pages} {64} (\bibinfo {year}
  {2018})}\BibitemShut {NoStop}%
\bibitem [{\citenamefont {Fern{\'a}ndez}\ \emph {et~al.}(2019)\citenamefont
  {Fern{\'a}ndez}, \citenamefont {Tchekhovskoy}, \citenamefont {Quataert},
  \citenamefont {Foucart},\ and\ \citenamefont {Kasen}}]{fernandez2019long}%
  \BibitemOpen
  \bibfield  {author} {\bibinfo {author} {\bibfnamefont {R.}~\bibnamefont
  {Fern{\'a}ndez}}, \bibinfo {author} {\bibfnamefont {A.}~\bibnamefont
  {Tchekhovskoy}}, \bibinfo {author} {\bibfnamefont {E.}~\bibnamefont
  {Quataert}}, \bibinfo {author} {\bibfnamefont {F.}~\bibnamefont {Foucart}},\
  and\ \bibinfo {author} {\bibfnamefont {D.}~\bibnamefont {Kasen}},\ }\bibfield
   {title} {\bibinfo {title} {Long-term grmhd simulations of neutron star
  merger accretion discs: implications for electromagnetic counterparts},\
  }\href@noop {} {\bibfield  {journal} {\bibinfo  {journal} {Monthly Notices of
  the Royal Astronomical Society}\ }\textbf {\bibinfo {volume} {482}},\
  \bibinfo {pages} {3373} (\bibinfo {year} {2019})}\BibitemShut {NoStop}%
\bibitem [{\citenamefont {Baumgarte}\ \emph {et~al.}(1999)\citenamefont
  {Baumgarte}, \citenamefont {Shapiro},\ and\ \citenamefont
  {Shibata}}]{baumgarte1999maximum}%
  \BibitemOpen
  \bibfield  {author} {\bibinfo {author} {\bibfnamefont {T.~W.}\ \bibnamefont
  {Baumgarte}}, \bibinfo {author} {\bibfnamefont {S.~L.}\ \bibnamefont
  {Shapiro}},\ and\ \bibinfo {author} {\bibfnamefont {M.}~\bibnamefont
  {Shibata}},\ }\bibfield  {title} {\bibinfo {title} {On the maximum mass of
  differentially rotating neutron stars},\ }\href@noop {} {\bibfield  {journal}
  {\bibinfo  {journal} {The Astrophysical Journal}\ }\textbf {\bibinfo {volume}
  {528}},\ \bibinfo {pages} {L29} (\bibinfo {year} {1999})}\BibitemShut
  {NoStop}%
\bibitem [{\citenamefont {Demorest}\ \emph {et~al.}(2010)\citenamefont
  {Demorest}, \citenamefont {Pennucci}, \citenamefont {Ransom}, \citenamefont
  {Roberts},\ and\ \citenamefont {Hessels}}]{demorest2010two}%
  \BibitemOpen
  \bibfield  {author} {\bibinfo {author} {\bibfnamefont {P.~B.}\ \bibnamefont
  {Demorest}}, \bibinfo {author} {\bibfnamefont {T.}~\bibnamefont {Pennucci}},
  \bibinfo {author} {\bibfnamefont {S.}~\bibnamefont {Ransom}}, \bibinfo
  {author} {\bibfnamefont {M.}~\bibnamefont {Roberts}},\ and\ \bibinfo {author}
  {\bibfnamefont {J.}~\bibnamefont {Hessels}},\ }\bibfield  {title} {\bibinfo
  {title} {A two-solar-mass neutron star measured using shapiro delay},\
  }\href@noop {} {\bibfield  {journal} {\bibinfo  {journal} {nature}\ }\textbf
  {\bibinfo {volume} {467}},\ \bibinfo {pages} {1081} (\bibinfo {year}
  {2010})}\BibitemShut {NoStop}%
\bibitem [{\citenamefont {Antoniadis}\ \emph {et~al.}(2013)\citenamefont
  {Antoniadis}, \citenamefont {Freire}, \citenamefont {Wex}, \citenamefont
  {Tauris}, \citenamefont {Lynch}, \citenamefont {Van~Kerkwijk}, \citenamefont
  {Kramer}, \citenamefont {Bassa}, \citenamefont {Dhillon}, \citenamefont
  {Driebe} \emph {et~al.}}]{antoniadis2013massive}%
  \BibitemOpen
  \bibfield  {author} {\bibinfo {author} {\bibfnamefont {J.}~\bibnamefont
  {Antoniadis}}, \bibinfo {author} {\bibfnamefont {P.~C.}\ \bibnamefont
  {Freire}}, \bibinfo {author} {\bibfnamefont {N.}~\bibnamefont {Wex}},
  \bibinfo {author} {\bibfnamefont {T.~M.}\ \bibnamefont {Tauris}}, \bibinfo
  {author} {\bibfnamefont {R.~S.}\ \bibnamefont {Lynch}}, \bibinfo {author}
  {\bibfnamefont {M.~H.}\ \bibnamefont {Van~Kerkwijk}}, \bibinfo {author}
  {\bibfnamefont {M.}~\bibnamefont {Kramer}}, \bibinfo {author} {\bibfnamefont
  {C.}~\bibnamefont {Bassa}}, \bibinfo {author} {\bibfnamefont {V.~S.}\
  \bibnamefont {Dhillon}}, \bibinfo {author} {\bibfnamefont {T.}~\bibnamefont
  {Driebe}}, \emph {et~al.},\ }\bibfield  {title} {\bibinfo {title} {A massive
  pulsar in a compact relativistic binary},\ }\href@noop {} {\bibfield
  {journal} {\bibinfo  {journal} {Science}\ }\textbf {\bibinfo {volume}
  {340}},\ \bibinfo {pages} {1233232} (\bibinfo {year} {2013})}\BibitemShut
  {NoStop}%
\bibitem [{\citenamefont {Shibata}\ \emph {et~al.}(2005)\citenamefont
  {Shibata}, \citenamefont {Taniguchi},\ and\ \citenamefont
  {Ury{\=u}}}]{shibata2005merger}%
  \BibitemOpen
  \bibfield  {author} {\bibinfo {author} {\bibfnamefont {M.}~\bibnamefont
  {Shibata}}, \bibinfo {author} {\bibfnamefont {K.}~\bibnamefont {Taniguchi}},\
  and\ \bibinfo {author} {\bibfnamefont {K.}~\bibnamefont {Ury{\=u}}},\
  }\bibfield  {title} {\bibinfo {title} {Merger of binary neutron stars with
  realistic equations of state in full general relativity},\ }\href@noop {}
  {\bibfield  {journal} {\bibinfo  {journal} {Physical Review D}\ }\textbf
  {\bibinfo {volume} {71}},\ \bibinfo {pages} {084021} (\bibinfo {year}
  {2005})}\BibitemShut {NoStop}%
\bibitem [{\citenamefont {Shibata}\ and\ \citenamefont
  {Taniguchi}(2006)}]{shibata2006merger}%
  \BibitemOpen
  \bibfield  {author} {\bibinfo {author} {\bibfnamefont {M.}~\bibnamefont
  {Shibata}}\ and\ \bibinfo {author} {\bibfnamefont {K.}~\bibnamefont
  {Taniguchi}},\ }\bibfield  {title} {\bibinfo {title} {Merger of binary
  neutron stars to a black hole: Disk mass, short gamma-ray bursts, and
  quasinormal mode ringing},\ }\href@noop {} {\bibfield  {journal} {\bibinfo
  {journal} {Physical Review D}\ }\textbf {\bibinfo {volume} {73}},\ \bibinfo
  {pages} {064027} (\bibinfo {year} {2006})}\BibitemShut {NoStop}%
\bibitem [{\citenamefont {Kiuchi}\ \emph {et~al.}(2009)\citenamefont {Kiuchi},
  \citenamefont {Sekiguchi}, \citenamefont {Shibata},\ and\ \citenamefont
  {Taniguchi}}]{kiuchi2009long}%
  \BibitemOpen
  \bibfield  {author} {\bibinfo {author} {\bibfnamefont {K.}~\bibnamefont
  {Kiuchi}}, \bibinfo {author} {\bibfnamefont {Y.}~\bibnamefont {Sekiguchi}},
  \bibinfo {author} {\bibfnamefont {M.}~\bibnamefont {Shibata}},\ and\ \bibinfo
  {author} {\bibfnamefont {K.}~\bibnamefont {Taniguchi}},\ }\bibfield  {title}
  {\bibinfo {title} {Long-term general relativistic simulation of binary
  neutron stars collapsing to a black hole},\ }\href@noop {} {\bibfield
  {journal} {\bibinfo  {journal} {Physical Review D}\ }\textbf {\bibinfo
  {volume} {80}},\ \bibinfo {pages} {064037} (\bibinfo {year}
  {2009})}\BibitemShut {NoStop}%
\bibitem [{\citenamefont {Hotokezaka}\ \emph {et~al.}(2011)\citenamefont
  {Hotokezaka}, \citenamefont {Kyutoku}, \citenamefont {Okawa}, \citenamefont
  {Shibata},\ and\ \citenamefont {Kiuchi}}]{hotokezaka2011binary}%
  \BibitemOpen
  \bibfield  {author} {\bibinfo {author} {\bibfnamefont {K.}~\bibnamefont
  {Hotokezaka}}, \bibinfo {author} {\bibfnamefont {K.}~\bibnamefont {Kyutoku}},
  \bibinfo {author} {\bibfnamefont {H.}~\bibnamefont {Okawa}}, \bibinfo
  {author} {\bibfnamefont {M.}~\bibnamefont {Shibata}},\ and\ \bibinfo {author}
  {\bibfnamefont {K.}~\bibnamefont {Kiuchi}},\ }\bibfield  {title} {\bibinfo
  {title} {Binary neutron star mergers: dependence on the nuclear equation of
  state},\ }\href@noop {} {\bibfield  {journal} {\bibinfo  {journal} {Physical
  Review D}\ }\textbf {\bibinfo {volume} {83}},\ \bibinfo {pages} {124008}
  (\bibinfo {year} {2011})}\BibitemShut {NoStop}%
\bibitem [{\citenamefont {Hotokezaka}\ \emph {et~al.}(2013)\citenamefont
  {Hotokezaka}, \citenamefont {Kiuchi}, \citenamefont {Kyutoku}, \citenamefont
  {Muranushi}, \citenamefont {Sekiguchi}, \citenamefont {Shibata},\ and\
  \citenamefont {Taniguchi}}]{hotokezaka2013remnant}%
  \BibitemOpen
  \bibfield  {author} {\bibinfo {author} {\bibfnamefont {K.}~\bibnamefont
  {Hotokezaka}}, \bibinfo {author} {\bibfnamefont {K.}~\bibnamefont {Kiuchi}},
  \bibinfo {author} {\bibfnamefont {K.}~\bibnamefont {Kyutoku}}, \bibinfo
  {author} {\bibfnamefont {T.}~\bibnamefont {Muranushi}}, \bibinfo {author}
  {\bibfnamefont {Y.-i.}\ \bibnamefont {Sekiguchi}}, \bibinfo {author}
  {\bibfnamefont {M.}~\bibnamefont {Shibata}},\ and\ \bibinfo {author}
  {\bibfnamefont {K.}~\bibnamefont {Taniguchi}},\ }\bibfield  {title} {\bibinfo
  {title} {Remnant massive neutron stars of binary neutron star mergers:
  Evolution process and gravitational waveform},\ }\href@noop {} {\bibfield
  {journal} {\bibinfo  {journal} {Physical Review D}\ }\textbf {\bibinfo
  {volume} {88}},\ \bibinfo {pages} {044026} (\bibinfo {year}
  {2013})}\BibitemShut {NoStop}%
\bibitem [{\citenamefont {Takami}\ \emph {et~al.}(2015)\citenamefont {Takami},
  \citenamefont {Rezzolla},\ and\ \citenamefont
  {Baiotti}}]{takami2015spectral}%
  \BibitemOpen
  \bibfield  {author} {\bibinfo {author} {\bibfnamefont {K.}~\bibnamefont
  {Takami}}, \bibinfo {author} {\bibfnamefont {L.}~\bibnamefont {Rezzolla}},\
  and\ \bibinfo {author} {\bibfnamefont {L.}~\bibnamefont {Baiotti}},\
  }\bibfield  {title} {\bibinfo {title} {Spectral properties of the post-merger
  gravitational-wave signal from binary neutron stars},\ }\href@noop {}
  {\bibfield  {journal} {\bibinfo  {journal} {Physical Review D}\ }\textbf
  {\bibinfo {volume} {91}},\ \bibinfo {pages} {064001} (\bibinfo {year}
  {2015})}\BibitemShut {NoStop}%
\bibitem [{\citenamefont {Dietrich}\ \emph {et~al.}(2015)\citenamefont
  {Dietrich}, \citenamefont {Bernuzzi}, \citenamefont {Ujevic},\ and\
  \citenamefont {Br{\"u}gmann}}]{dietrich2015numerical}%
  \BibitemOpen
  \bibfield  {author} {\bibinfo {author} {\bibfnamefont {T.}~\bibnamefont
  {Dietrich}}, \bibinfo {author} {\bibfnamefont {S.}~\bibnamefont {Bernuzzi}},
  \bibinfo {author} {\bibfnamefont {M.}~\bibnamefont {Ujevic}},\ and\ \bibinfo
  {author} {\bibfnamefont {B.}~\bibnamefont {Br{\"u}gmann}},\ }\bibfield
  {title} {\bibinfo {title} {Numerical relativity simulations of neutron star
  merger remnants using conservative mesh refinement},\ }\href@noop {}
  {\bibfield  {journal} {\bibinfo  {journal} {Physical Review D}\ }\textbf
  {\bibinfo {volume} {91}},\ \bibinfo {pages} {124041} (\bibinfo {year}
  {2015})}\BibitemShut {NoStop}%
\bibitem [{\citenamefont {Bernuzzi}\ \emph {et~al.}(2016)\citenamefont
  {Bernuzzi}, \citenamefont {Radice}, \citenamefont {Ott}, \citenamefont
  {Roberts}, \citenamefont {M{\"o}sta},\ and\ \citenamefont
  {Galeazzi}}]{bernuzzi2016loud}%
  \BibitemOpen
  \bibfield  {author} {\bibinfo {author} {\bibfnamefont {S.}~\bibnamefont
  {Bernuzzi}}, \bibinfo {author} {\bibfnamefont {D.}~\bibnamefont {Radice}},
  \bibinfo {author} {\bibfnamefont {C.~D.}\ \bibnamefont {Ott}}, \bibinfo
  {author} {\bibfnamefont {L.~F.}\ \bibnamefont {Roberts}}, \bibinfo {author}
  {\bibfnamefont {P.}~\bibnamefont {M{\"o}sta}},\ and\ \bibinfo {author}
  {\bibfnamefont {F.}~\bibnamefont {Galeazzi}},\ }\bibfield  {title} {\bibinfo
  {title} {How loud are neutron star mergers?},\ }\href@noop {} {\bibfield
  {journal} {\bibinfo  {journal} {Physical Review D}\ }\textbf {\bibinfo
  {volume} {94}},\ \bibinfo {pages} {024023} (\bibinfo {year}
  {2016})}\BibitemShut {NoStop}%
\bibitem [{\citenamefont {Ciolfi}\ \emph {et~al.}(2017)\citenamefont {Ciolfi},
  \citenamefont {Kastaun}, \citenamefont {Giacomazzo}, \citenamefont
  {Endrizzi}, \citenamefont {Siegel},\ and\ \citenamefont
  {Perna}}]{ciolfi2017general}%
  \BibitemOpen
  \bibfield  {author} {\bibinfo {author} {\bibfnamefont {R.}~\bibnamefont
  {Ciolfi}}, \bibinfo {author} {\bibfnamefont {W.}~\bibnamefont {Kastaun}},
  \bibinfo {author} {\bibfnamefont {B.}~\bibnamefont {Giacomazzo}}, \bibinfo
  {author} {\bibfnamefont {A.}~\bibnamefont {Endrizzi}}, \bibinfo {author}
  {\bibfnamefont {D.~M.}\ \bibnamefont {Siegel}},\ and\ \bibinfo {author}
  {\bibfnamefont {R.}~\bibnamefont {Perna}},\ }\bibfield  {title} {\bibinfo
  {title} {General relativistic magnetohydrodynamic simulations of binary
  neutron star mergers forming a long-lived neutron star},\ }\href@noop {}
  {\bibfield  {journal} {\bibinfo  {journal} {Physical Review D}\ }\textbf
  {\bibinfo {volume} {95}},\ \bibinfo {pages} {063016} (\bibinfo {year}
  {2017})}\BibitemShut {NoStop}%
\bibitem [{\citenamefont {Cumming}\ \emph {et~al.}(2017)\citenamefont
  {Cumming}, \citenamefont {Brown}, \citenamefont {Fattoyev}, \citenamefont
  {Horowitz}, \citenamefont {Page},\ and\ \citenamefont
  {Reddy}}]{cumming2017lower}%
  \BibitemOpen
  \bibfield  {author} {\bibinfo {author} {\bibfnamefont {A.}~\bibnamefont
  {Cumming}}, \bibinfo {author} {\bibfnamefont {E.~F.}\ \bibnamefont {Brown}},
  \bibinfo {author} {\bibfnamefont {F.~J.}\ \bibnamefont {Fattoyev}}, \bibinfo
  {author} {\bibfnamefont {C.}~\bibnamefont {Horowitz}}, \bibinfo {author}
  {\bibfnamefont {D.}~\bibnamefont {Page}},\ and\ \bibinfo {author}
  {\bibfnamefont {S.}~\bibnamefont {Reddy}},\ }\bibfield  {title} {\bibinfo
  {title} {Lower limit on the heat capacity of the neutron star core},\
  }\href@noop {} {\bibfield  {journal} {\bibinfo  {journal} {Physical Review
  C}\ }\textbf {\bibinfo {volume} {95}},\ \bibinfo {pages} {025806} (\bibinfo
  {year} {2017})}\BibitemShut {NoStop}%
\bibitem [{\citenamefont {Janka}\ and\ \citenamefont
  {Ruffert}(1995)}]{janka1995can}%
  \BibitemOpen
  \bibfield  {author} {\bibinfo {author} {\bibfnamefont {H.-T.}\ \bibnamefont
  {Janka}}\ and\ \bibinfo {author} {\bibfnamefont {M.}~\bibnamefont
  {Ruffert}},\ }\bibfield  {title} {\bibinfo {title} {Can neutrinos from
  neutron star mergers power gamma-ray bursts?},\ }\href@noop {} {\bibfield
  {journal} {\bibinfo  {journal} {arXiv preprint astro-ph/9512144}\ } (\bibinfo
  {year} {1995})}\BibitemShut {NoStop}%
\bibitem [{\citenamefont {Kochanek}(1992)}]{kochanek1992coalescing}%
  \BibitemOpen
  \bibfield  {author} {\bibinfo {author} {\bibfnamefont {C.~S.}\ \bibnamefont
  {Kochanek}},\ }\bibfield  {title} {\bibinfo {title} {Coalescing binary
  neutron stars},\ }\href@noop {} {\bibfield  {journal} {\bibinfo  {journal}
  {Astrophysical Journal, Part 1 (ISSN 0004-637X), vol. 398, no. 1, p.
  234-247.}\ }\textbf {\bibinfo {volume} {398}},\ \bibinfo {pages} {234}
  (\bibinfo {year} {1992})}\BibitemShut {NoStop}%
\bibitem [{\citenamefont {Bildsten}\ and\ \citenamefont
  {Cutler}(1992)}]{bildsten1992tidal}%
  \BibitemOpen
  \bibfield  {author} {\bibinfo {author} {\bibfnamefont {L.}~\bibnamefont
  {Bildsten}}\ and\ \bibinfo {author} {\bibfnamefont {C.}~\bibnamefont
  {Cutler}},\ }\bibfield  {title} {\bibinfo {title} {Tidal interactions of
  inspiraling compact binaries},\ }\href@noop {} {\bibfield  {journal}
  {\bibinfo  {journal} {Astrophysical Journal, Part 1 (ISSN 0004-637X), vol.
  400, no. 1, p. 175-180.}\ }\textbf {\bibinfo {volume} {400}},\ \bibinfo
  {pages} {175} (\bibinfo {year} {1992})}\BibitemShut {NoStop}%
\bibitem [{\citenamefont {Flowers}\ and\ \citenamefont
  {Itoh}(1979)}]{flowers1979transport}%
  \BibitemOpen
  \bibfield  {author} {\bibinfo {author} {\bibfnamefont {E.}~\bibnamefont
  {Flowers}}\ and\ \bibinfo {author} {\bibfnamefont {N.}~\bibnamefont {Itoh}},\
  }\bibfield  {title} {\bibinfo {title} {Transport properties of dense matter.
  ii},\ }\href@noop {} {\bibfield  {journal} {\bibinfo  {journal}
  {Astrophysical Journal, Part 1, vol. 230, June 15, 1979, p. 847-858.}\
  }\textbf {\bibinfo {volume} {230}},\ \bibinfo {pages} {847} (\bibinfo {year}
  {1979})}\BibitemShut {NoStop}%
\bibitem [{\citenamefont {Orsaria}\ \emph {et~al.}(2019)\citenamefont
  {Orsaria}, \citenamefont {Malfatti}, \citenamefont {Mariani}, \citenamefont
  {Ranea-Sandoval}, \citenamefont {Garc{\'\i}a}, \citenamefont {Spinella},
  \citenamefont {Contrera}, \citenamefont {Lugones},\ and\ \citenamefont
  {Weber}}]{orsaria2019phase}%
  \BibitemOpen
  \bibfield  {author} {\bibinfo {author} {\bibfnamefont {M.~G.}\ \bibnamefont
  {Orsaria}}, \bibinfo {author} {\bibfnamefont {G.}~\bibnamefont {Malfatti}},
  \bibinfo {author} {\bibfnamefont {M.}~\bibnamefont {Mariani}}, \bibinfo
  {author} {\bibfnamefont {I.~F.}\ \bibnamefont {Ranea-Sandoval}}, \bibinfo
  {author} {\bibfnamefont {F.}~\bibnamefont {Garc{\'\i}a}}, \bibinfo {author}
  {\bibfnamefont {W.~M.}\ \bibnamefont {Spinella}}, \bibinfo {author}
  {\bibfnamefont {G.~A.}\ \bibnamefont {Contrera}}, \bibinfo {author}
  {\bibfnamefont {G.}~\bibnamefont {Lugones}},\ and\ \bibinfo {author}
  {\bibfnamefont {F.}~\bibnamefont {Weber}},\ }\bibfield  {title} {\bibinfo
  {title} {Phase transitions in neutron stars and their links to gravitational
  waves},\ }\href@noop {} {\bibfield  {journal} {\bibinfo  {journal} {Journal
  of Physics G: Nuclear and Particle Physics}\ }\textbf {\bibinfo {volume}
  {46}},\ \bibinfo {pages} {073002} (\bibinfo {year} {2019})}\BibitemShut
  {NoStop}%
\bibitem [{\citenamefont {Cutler}\ and\ \citenamefont
  {Thorne}(2013)}]{Cutler:2002me}%
  \BibitemOpen
  \bibfield  {author} {\bibinfo {author} {\bibfnamefont {C.}~\bibnamefont
  {Cutler}}\ and\ \bibinfo {author} {\bibfnamefont {K.~S.}\ \bibnamefont
  {Thorne}},\ }\bibfield  {title} {\bibinfo {title} {{An Overview of
  gravitational wave sources}},\ }in\ \href
  {https://doi.org/10.1142/9789812776556_0004} {\emph {\bibinfo {booktitle}
  {{16th International Conference on General Relativity and Gravitation
  (GR16)}}}}\ (\bibinfo {year} {2013})\ pp.\ \bibinfo {pages} {72--111},\
  \Eprint {https://arxiv.org/abs/gr-qc/0204090} {arXiv:gr-qc/0204090}
  \BibitemShut {NoStop}%
\bibitem [{\citenamefont {Rozwadowska}\ \emph {et~al.}(2021)\citenamefont
  {Rozwadowska}, \citenamefont {Vissani},\ and\ \citenamefont
  {Cappellaro}}]{Rozwadowska:2020nab}%
  \BibitemOpen
  \bibfield  {author} {\bibinfo {author} {\bibfnamefont {K.}~\bibnamefont
  {Rozwadowska}}, \bibinfo {author} {\bibfnamefont {F.}~\bibnamefont
  {Vissani}},\ and\ \bibinfo {author} {\bibfnamefont {E.}~\bibnamefont
  {Cappellaro}},\ }\bibfield  {title} {\bibinfo {title} {{On the rate of core
  collapse supernovae in the milky way}},\ }\href
  {https://doi.org/10.1016/j.newast.2020.101498} {\bibfield  {journal}
  {\bibinfo  {journal} {New Astron.}\ }\textbf {\bibinfo {volume} {83}},\
  \bibinfo {pages} {101498} (\bibinfo {year} {2021})},\ \Eprint
  {https://arxiv.org/abs/2009.03438} {arXiv:2009.03438 [astro-ph.HE]}
  \BibitemShut {NoStop}%
\bibitem [{\citenamefont {Fryer}\ \emph {et~al.}(2023)\citenamefont {Fryer},
  \citenamefont {Burns}, \citenamefont {Hungerford}, \citenamefont {Safi-Harb},
  \citenamefont {Wollaeger}, \citenamefont {Miller}, \citenamefont {Negro},
  \citenamefont {Anandagoda},\ and\ \citenamefont {Hartmann}}]{Fryer:2023ehc}%
  \BibitemOpen
  \bibfield  {author} {\bibinfo {author} {\bibfnamefont {C.~L.}\ \bibnamefont
  {Fryer}}, \bibinfo {author} {\bibfnamefont {E.}~\bibnamefont {Burns}},
  \bibinfo {author} {\bibfnamefont {A.}~\bibnamefont {Hungerford}}, \bibinfo
  {author} {\bibfnamefont {S.}~\bibnamefont {Safi-Harb}}, \bibinfo {author}
  {\bibfnamefont {R.~T.}\ \bibnamefont {Wollaeger}}, \bibinfo {author}
  {\bibfnamefont {R.~S.}\ \bibnamefont {Miller}}, \bibinfo {author}
  {\bibfnamefont {M.}~\bibnamefont {Negro}}, \bibinfo {author} {\bibfnamefont
  {S.}~\bibnamefont {Anandagoda}},\ and\ \bibinfo {author} {\bibfnamefont
  {D.~H.}\ \bibnamefont {Hartmann}},\ }\bibfield  {title} {\bibinfo {title}
  {{Multimessenger Diagnostics of the Engine behind Core-collapse
  Supernovae}},\ }\href {https://doi.org/10.3847/1538-4357/ace0c3} {\bibfield
  {journal} {\bibinfo  {journal} {Astrophys. J.}\ }\textbf {\bibinfo {volume}
  {956}},\ \bibinfo {pages} {19} (\bibinfo {year} {2023})},\ \Eprint
  {https://arxiv.org/abs/2305.06134} {arXiv:2305.06134 [astro-ph.HE]}
  \BibitemShut {NoStop}%
\bibitem [{\citenamefont {Di~Palma}\ \emph {et~al.}(2023)\citenamefont
  {Di~Palma}, \citenamefont {Cerd\'a-Dur\'an}, \citenamefont {Drago},
  \citenamefont {L\'opez}, \citenamefont {Ricci},\ and\ \citenamefont
  {Veutro}}]{DiPalma:2023hxs}%
  \BibitemOpen
  \bibfield  {author} {\bibinfo {author} {\bibfnamefont {I.}~\bibnamefont
  {Di~Palma}}, \bibinfo {author} {\bibfnamefont {P.}~\bibnamefont
  {Cerd\'a-Dur\'an}}, \bibinfo {author} {\bibfnamefont {M.}~\bibnamefont
  {Drago}}, \bibinfo {author} {\bibfnamefont {M.}~\bibnamefont {L\'opez}},
  \bibinfo {author} {\bibfnamefont {F.}~\bibnamefont {Ricci}},\ and\ \bibinfo
  {author} {\bibfnamefont {A.}~\bibnamefont {Veutro}},\ }\bibfield  {title}
  {\bibinfo {title} {{Multimessenger challenges for the detection of core
  collapse supernovae}},\ }\href {https://doi.org/10.22323/1.444.1573}
  {\bibfield  {journal} {\bibinfo  {journal} {PoS}\ }\textbf {\bibinfo {volume}
  {ICRC2023}},\ \bibinfo {pages} {1573} (\bibinfo {year} {2023})}\BibitemShut
  {NoStop}%
\bibitem [{\citenamefont {M\"uller}(2020)}]{Muller:2020ard}%
  \BibitemOpen
  \bibfield  {author} {\bibinfo {author} {\bibfnamefont {B.}~\bibnamefont
  {M\"uller}},\ }\bibfield  {title} {\bibinfo {title} {{Hydrodynamics of
  core-collapse supernovae and their progenitors}},\ }\href
  {https://doi.org/10.1007/s41115-020-0008-5} {\bibfield  {journal} {\bibinfo
  {journal} {Astrophysics}\ }\textbf {\bibinfo {volume} {6}},\ \bibinfo {pages}
  {3} (\bibinfo {year} {2020})},\ \Eprint {https://arxiv.org/abs/2006.05083}
  {arXiv:2006.05083 [astro-ph.SR]} \BibitemShut {NoStop}%
\bibitem [{\citenamefont {Abdikamalov}\ \emph {et~al.}(2020)\citenamefont
  {Abdikamalov}, \citenamefont {Pagliaroli},\ and\ \citenamefont
  {Radice}}]{Abdikamalov:2020jzn}%
  \BibitemOpen
  \bibfield  {author} {\bibinfo {author} {\bibfnamefont {E.}~\bibnamefont
  {Abdikamalov}}, \bibinfo {author} {\bibfnamefont {G.}~\bibnamefont
  {Pagliaroli}},\ and\ \bibinfo {author} {\bibfnamefont {D.}~\bibnamefont
  {Radice}},\ }\href@noop {} {\bibinfo {title} {{Gravitational Waves from
  Core-Collapse Supernovae}}} (\bibinfo {year} {2020}),\ \Eprint
  {https://arxiv.org/abs/2010.04356} {arXiv:2010.04356 [astro-ph.SR]}
  \BibitemShut {NoStop}%
\bibitem [{\citenamefont {{Heger}}\ \emph {et~al.}(2003)\citenamefont
  {{Heger}}, \citenamefont {{Fryer}}, \citenamefont {{Woosley}}, \citenamefont
  {{Langer}},\ and\ \citenamefont {{Hartmann}}}]{Heger03}%
  \BibitemOpen
  \bibfield  {author} {\bibinfo {author} {\bibfnamefont {A.}~\bibnamefont
  {{Heger}}}, \bibinfo {author} {\bibfnamefont {C.~L.}\ \bibnamefont
  {{Fryer}}}, \bibinfo {author} {\bibfnamefont {S.~E.}\ \bibnamefont
  {{Woosley}}}, \bibinfo {author} {\bibfnamefont {N.}~\bibnamefont
  {{Langer}}},\ and\ \bibinfo {author} {\bibfnamefont {D.~H.}\ \bibnamefont
  {{Hartmann}}},\ }\bibfield  {title} {\bibinfo {title} {{How Massive Single
  Stars End Their Life}},\ }\href {https://doi.org/10.1086/375341} {\bibfield
  {journal} {\bibinfo  {journal} {Astrophys. J.}\ }\textbf {\bibinfo {volume}
  {591}},\ \bibinfo {pages} {288} (\bibinfo {year} {2003})},\ \Eprint
  {https://arxiv.org/abs/arXiv:astro-ph/0212469} {arXiv:astro-ph/0212469}
  \BibitemShut {NoStop}%
\bibitem [{\citenamefont {{Woosley}}\ \emph {et~al.}(2002)\citenamefont
  {{Woosley}}, \citenamefont {{Heger}},\ and\ \citenamefont
  {{Weaver}}}]{Woosley02}%
  \BibitemOpen
  \bibfield  {author} {\bibinfo {author} {\bibfnamefont {S.~E.}\ \bibnamefont
  {{Woosley}}}, \bibinfo {author} {\bibfnamefont {A.}~\bibnamefont {{Heger}}},\
  and\ \bibinfo {author} {\bibfnamefont {T.~A.}\ \bibnamefont {{Weaver}}},\
  }\bibfield  {title} {\bibinfo {title} {{The evolution and explosion of
  massive stars}},\ }\href@noop {} {\bibfield  {journal} {\bibinfo  {journal}
  {Rev. Mod. Phys.}\ }\textbf {\bibinfo {volume} {74}},\ \bibinfo {pages}
  {1015} (\bibinfo {year} {2002})}\BibitemShut {NoStop}%
\bibitem [{\citenamefont {{Woosley}}\ and\ \citenamefont
  {{Heger}}(2006)}]{Woosley06}%
  \BibitemOpen
  \bibfield  {author} {\bibinfo {author} {\bibfnamefont {S.~E.}\ \bibnamefont
  {{Woosley}}}\ and\ \bibinfo {author} {\bibfnamefont {A.}~\bibnamefont
  {{Heger}}},\ }\bibfield  {title} {\bibinfo {title} {{The Progenitor Stars of
  Gamma-Ray Bursts}},\ }\href {https://doi.org/10.1086/498500} {\bibfield
  {journal} {\bibinfo  {journal} {Astrophys. J.}\ }\textbf {\bibinfo {volume}
  {637}},\ \bibinfo {pages} {914} (\bibinfo {year} {2006})},\ \Eprint
  {https://arxiv.org/abs/arXiv:astro-ph/0508175} {arXiv:astro-ph/0508175}
  \BibitemShut {NoStop}%
\bibitem [{\citenamefont {Woosley}\ and\ \citenamefont
  {Janka}(2005)}]{Woosley:2006ie}%
  \BibitemOpen
  \bibfield  {author} {\bibinfo {author} {\bibfnamefont {S.}~\bibnamefont
  {Woosley}}\ and\ \bibinfo {author} {\bibfnamefont {T.}~\bibnamefont
  {Janka}},\ }\bibfield  {title} {\bibinfo {title} {{The physics of
  core-collapse supernovae}},\ }\href {https://doi.org/10.1038/nphys172}
  {\bibfield  {journal} {\bibinfo  {journal} {Nature Phys.}\ }\textbf {\bibinfo
  {volume} {1}},\ \bibinfo {pages} {147} (\bibinfo {year} {2005})},\ \Eprint
  {https://arxiv.org/abs/astro-ph/0601261} {arXiv:astro-ph/0601261}
  \BibitemShut {NoStop}%
\bibitem [{\citenamefont {Andresen}\ \emph {et~al.}(2017)\citenamefont
  {Andresen}, \citenamefont {M\"uller}, \citenamefont {M\"uller},\ and\
  \citenamefont {Janka}}]{Andresen:2016pdt}%
  \BibitemOpen
  \bibfield  {author} {\bibinfo {author} {\bibfnamefont {H.}~\bibnamefont
  {Andresen}}, \bibinfo {author} {\bibfnamefont {B.}~\bibnamefont {M\"uller}},
  \bibinfo {author} {\bibfnamefont {E.}~\bibnamefont {M\"uller}},\ and\
  \bibinfo {author} {\bibfnamefont {H.-T.}\ \bibnamefont {Janka}},\ }\bibfield
  {title} {\bibinfo {title} {{Gravitational Wave Signals from 3D Neutrino
  Hydrodynamics Simulations of Core-Collapse Supernovae}},\ }\href
  {https://doi.org/10.1093/mnras/stx618} {\bibfield  {journal} {\bibinfo
  {journal} {Mon. Not. Roy. Astron. Soc.}\ }\textbf {\bibinfo {volume} {468}},\
  \bibinfo {pages} {2032} (\bibinfo {year} {2017})},\ \Eprint
  {https://arxiv.org/abs/1607.05199} {arXiv:1607.05199 [astro-ph.HE]}
  \BibitemShut {NoStop}%
\bibitem [{\citenamefont {Andresen}\ \emph {et~al.}(2019)\citenamefont
  {Andresen}, \citenamefont {M\"uller}, \citenamefont {Janka}, \citenamefont
  {Summa}, \citenamefont {Gill},\ and\ \citenamefont
  {Zanolin}}]{Andresen:2018aom}%
  \BibitemOpen
  \bibfield  {author} {\bibinfo {author} {\bibfnamefont {H.}~\bibnamefont
  {Andresen}}, \bibinfo {author} {\bibfnamefont {E.}~\bibnamefont {M\"uller}},
  \bibinfo {author} {\bibfnamefont {H.}~\bibnamefont {Janka}}, \bibinfo
  {author} {\bibfnamefont {A.}~\bibnamefont {Summa}}, \bibinfo {author}
  {\bibfnamefont {K.}~\bibnamefont {Gill}},\ and\ \bibinfo {author}
  {\bibfnamefont {M.}~\bibnamefont {Zanolin}},\ }\bibfield  {title} {\bibinfo
  {title} {{Gravitational waves from 3D core-collapse supernova models: The
  impact of moderate progenitor rotation}},\ }\href
  {https://doi.org/10.1093/mnras/stz990} {\bibfield  {journal} {\bibinfo
  {journal} {Mon. Not. Roy. Astron. Soc.}\ }\textbf {\bibinfo {volume} {486}},\
  \bibinfo {pages} {2238} (\bibinfo {year} {2019})},\ \Eprint
  {https://arxiv.org/abs/1810.07638} {arXiv:1810.07638 [astro-ph.HE]}
  \BibitemShut {NoStop}%
\bibitem [{\citenamefont {Radice}\ \emph {et~al.}(2019)\citenamefont {Radice},
  \citenamefont {Morozova}, \citenamefont {Burrows}, \citenamefont
  {Vartanyan},\ and\ \citenamefont {Nagakura}}]{Radice:2018usf}%
  \BibitemOpen
  \bibfield  {author} {\bibinfo {author} {\bibfnamefont {D.}~\bibnamefont
  {Radice}}, \bibinfo {author} {\bibfnamefont {V.}~\bibnamefont {Morozova}},
  \bibinfo {author} {\bibfnamefont {A.}~\bibnamefont {Burrows}}, \bibinfo
  {author} {\bibfnamefont {D.}~\bibnamefont {Vartanyan}},\ and\ \bibinfo
  {author} {\bibfnamefont {H.}~\bibnamefont {Nagakura}},\ }\bibfield  {title}
  {\bibinfo {title} {{Characterizing the Gravitational Wave Signal from
  Core-Collapse Supernovae}},\ }\href
  {https://doi.org/10.3847/2041-8213/ab191a} {\bibfield  {journal} {\bibinfo
  {journal} {Astrophys. J. Lett.}\ }\textbf {\bibinfo {volume} {876}},\
  \bibinfo {pages} {L9} (\bibinfo {year} {2019})},\ \Eprint
  {https://arxiv.org/abs/1812.07703} {arXiv:1812.07703 [astro-ph.HE]}
  \BibitemShut {NoStop}%
\bibitem [{\citenamefont {Hsieh}\ \emph {et~al.}(2023)\citenamefont {Hsieh},
  \citenamefont {Cabezón}, \citenamefont {Ma},\ and\ \citenamefont
  {Pan}}]{hsieh2023new}%
  \BibitemOpen
  \bibfield  {author} {\bibinfo {author} {\bibfnamefont {H.-F.}\ \bibnamefont
  {Hsieh}}, \bibinfo {author} {\bibfnamefont {R.}~\bibnamefont {Cabezón}},
  \bibinfo {author} {\bibfnamefont {L.-T.}\ \bibnamefont {Ma}},\ and\ \bibinfo
  {author} {\bibfnamefont {K.-C.}\ \bibnamefont {Pan}},\ }\href@noop {}
  {\bibinfo {title} {A new kilohertz gravitational-wave feature from rapidly
  rotating core-collapse supernovae}} (\bibinfo {year} {2023}),\ \Eprint
  {https://arxiv.org/abs/2310.20411} {arXiv:2310.20411 [astro-ph.HE]}
  \BibitemShut {NoStop}%
\bibitem [{\citenamefont {Roma}\ \emph {et~al.}(2019)\citenamefont {Roma},
  \citenamefont {Powell}, \citenamefont {Heng},\ and\ \citenamefont
  {Frey}}]{Roma:2019kcd}%
  \BibitemOpen
  \bibfield  {author} {\bibinfo {author} {\bibfnamefont {V.}~\bibnamefont
  {Roma}}, \bibinfo {author} {\bibfnamefont {J.}~\bibnamefont {Powell}},
  \bibinfo {author} {\bibfnamefont {I.~S.}\ \bibnamefont {Heng}},\ and\
  \bibinfo {author} {\bibfnamefont {R.}~\bibnamefont {Frey}},\ }\bibfield
  {title} {\bibinfo {title} {{Astrophysics with core-collapse supernova
  gravitational wave signals in the next generation of gravitational wave
  detectors}},\ }\href {https://doi.org/10.1103/PhysRevD.99.063018} {\bibfield
  {journal} {\bibinfo  {journal} {Phys. Rev. D}\ }\textbf {\bibinfo {volume}
  {99}},\ \bibinfo {pages} {063018} (\bibinfo {year} {2019})},\ \Eprint
  {https://arxiv.org/abs/1901.08692} {arXiv:1901.08692 [astro-ph.IM]}
  \BibitemShut {NoStop}%
\bibitem [{\citenamefont {Scheidegger}\ \emph {et~al.}(2010)\citenamefont
  {Scheidegger}, \citenamefont {Kaeppeli}, \citenamefont {Whitehouse},
  \citenamefont {Fischer},\ and\ \citenamefont
  {Liebendoerfer}}]{Scheidegger:2010en}%
  \BibitemOpen
  \bibfield  {author} {\bibinfo {author} {\bibfnamefont {S.}~\bibnamefont
  {Scheidegger}}, \bibinfo {author} {\bibfnamefont {R.}~\bibnamefont
  {Kaeppeli}}, \bibinfo {author} {\bibfnamefont {S.~C.}\ \bibnamefont
  {Whitehouse}}, \bibinfo {author} {\bibfnamefont {T.}~\bibnamefont
  {Fischer}},\ and\ \bibinfo {author} {\bibfnamefont {M.}~\bibnamefont
  {Liebendoerfer}},\ }\bibfield  {title} {\bibinfo {title} {{The Influence of
  Model Parameters on the Prediction of Gravitational wave Signals from Stellar
  Core Collapse}},\ }\href {https://doi.org/10.1051/0004-6361/200913220}
  {\bibfield  {journal} {\bibinfo  {journal} {Astron. Astrophys.}\ }\textbf
  {\bibinfo {volume} {514}},\ \bibinfo {pages} {A51} (\bibinfo {year}
  {2010})},\ \Eprint {https://arxiv.org/abs/1001.1570} {arXiv:1001.1570
  [astro-ph.HE]} \BibitemShut {NoStop}%
\bibitem [{\citenamefont {Muller}\ \emph {et~al.}(2012)\citenamefont {Muller},
  \citenamefont {Janka},\ and\ \citenamefont {Wongwathanarat}}]{Muller:2011yi}%
  \BibitemOpen
  \bibfield  {author} {\bibinfo {author} {\bibfnamefont {E.}~\bibnamefont
  {Muller}}, \bibinfo {author} {\bibfnamefont {H.~T.}\ \bibnamefont {Janka}},\
  and\ \bibinfo {author} {\bibfnamefont {A.}~\bibnamefont {Wongwathanarat}},\
  }\bibfield  {title} {\bibinfo {title} {{Parametrized 3D models of
  neutrino-driven supernova explosions: Neutrino emission asymmetries and
  gravitational-wave signals}},\ }\href
  {https://doi.org/10.1051/0004-6361/201117611} {\bibfield  {journal} {\bibinfo
   {journal} {Astron. Astrophys.}\ }\textbf {\bibinfo {volume} {537}},\
  \bibinfo {pages} {A63} (\bibinfo {year} {2012})},\ \Eprint
  {https://arxiv.org/abs/1106.6301} {arXiv:1106.6301 [astro-ph.SR]}
  \BibitemShut {NoStop}%
\bibitem [{\citenamefont {Kuroda}\ \emph {et~al.}(2016)\citenamefont {Kuroda},
  \citenamefont {Kotake},\ and\ \citenamefont {Takiwaki}}]{Kuroda:2016bjd}%
  \BibitemOpen
  \bibfield  {author} {\bibinfo {author} {\bibfnamefont {T.}~\bibnamefont
  {Kuroda}}, \bibinfo {author} {\bibfnamefont {K.}~\bibnamefont {Kotake}},\
  and\ \bibinfo {author} {\bibfnamefont {T.}~\bibnamefont {Takiwaki}},\
  }\bibfield  {title} {\bibinfo {title} {{A new Gravitational-wave Signature
  From Standing Accretion Shock Instability in Supernovae}},\ }\href
  {https://doi.org/10.3847/2041-8205/829/1/L14} {\bibfield  {journal} {\bibinfo
   {journal} {Astrophys. J. Lett.}\ }\textbf {\bibinfo {volume} {829}},\
  \bibinfo {pages} {L14} (\bibinfo {year} {2016})},\ \Eprint
  {https://arxiv.org/abs/1605.09215} {arXiv:1605.09215 [astro-ph.HE]}
  \BibitemShut {NoStop}%
\bibitem [{\citenamefont {Mezzacappa}\ \emph {et~al.}(2020)\citenamefont
  {Mezzacappa} \emph {et~al.}}]{Mezzacappa:2020lsn}%
  \BibitemOpen
  \bibfield  {author} {\bibinfo {author} {\bibfnamefont {A.}~\bibnamefont
  {Mezzacappa}} \emph {et~al.},\ }\bibfield  {title} {\bibinfo {title}
  {{Gravitational-wave signal of a core-collapse supernova explosion of a 15
  $M_{\odot}$ star}},\ }\href {https://doi.org/10.1103/PhysRevD.102.023027}
  {\bibfield  {journal} {\bibinfo  {journal} {Phys. Rev. D}\ }\textbf {\bibinfo
  {volume} {102}},\ \bibinfo {pages} {023027} (\bibinfo {year} {2020})},\
  \Eprint {https://arxiv.org/abs/2007.15099} {arXiv:2007.15099 [astro-ph.HE]}
  \BibitemShut {NoStop}%
\bibitem [{\citenamefont {Powell}\ and\ \citenamefont
  {M\"uller}(2019)}]{Powell:2018isq}%
  \BibitemOpen
  \bibfield  {author} {\bibinfo {author} {\bibfnamefont {J.}~\bibnamefont
  {Powell}}\ and\ \bibinfo {author} {\bibfnamefont {B.}~\bibnamefont
  {M\"uller}},\ }\bibfield  {title} {\bibinfo {title} {{Gravitational Wave
  Emission from 3D Explosion Models of Core-Collapse Supernovae with Low and
  Normal Explosion Energies}},\ }\href {https://doi.org/10.1093/mnras/stz1304}
  {\bibfield  {journal} {\bibinfo  {journal} {Mon. Not. Roy. Astron. Soc.}\
  }\textbf {\bibinfo {volume} {487}},\ \bibinfo {pages} {1178} (\bibinfo {year}
  {2019})},\ \Eprint {https://arxiv.org/abs/1812.05738} {arXiv:1812.05738
  [astro-ph.HE]} \BibitemShut {NoStop}%
\bibitem [{\citenamefont {Shibagaki}\ \emph {et~al.}(2021)\citenamefont
  {Shibagaki}, \citenamefont {Kuroda}, \citenamefont {Kotake},\ and\
  \citenamefont {Takiwaki}}]{Shibagaki:2020ksk}%
  \BibitemOpen
  \bibfield  {author} {\bibinfo {author} {\bibfnamefont {S.}~\bibnamefont
  {Shibagaki}}, \bibinfo {author} {\bibfnamefont {T.}~\bibnamefont {Kuroda}},
  \bibinfo {author} {\bibfnamefont {K.}~\bibnamefont {Kotake}},\ and\ \bibinfo
  {author} {\bibfnamefont {T.}~\bibnamefont {Takiwaki}},\ }\bibfield  {title}
  {\bibinfo {title} {{Characteristic Time Variability of Gravitational-Wave and
  Neutrino Signals from Three-dimensional Simulations of Non-Rotating and
  Rapidly Rotating Stellar Core-Collapse}},\ }\href
  {https://doi.org/10.1093/mnras/stab228} {\bibfield  {journal} {\bibinfo
  {journal} {Mon. Not. Roy. Astron. Soc.}\ }\textbf {\bibinfo {volume} {502}},\
  \bibinfo {pages} {3066} (\bibinfo {year} {2021})},\ \Eprint
  {https://arxiv.org/abs/2010.03882} {arXiv:2010.03882 [astro-ph.HE]}
  \BibitemShut {NoStop}%
\bibitem [{\citenamefont {Furusawa}\ and\ \citenamefont
  {Nagakura}(2023)}]{Furusawa:2022ktu}%
  \BibitemOpen
  \bibfield  {author} {\bibinfo {author} {\bibfnamefont {S.}~\bibnamefont
  {Furusawa}}\ and\ \bibinfo {author} {\bibfnamefont {H.}~\bibnamefont
  {Nagakura}},\ }\bibfield  {title} {\bibinfo {title} {{Nuclei in core-collapse
  supernovae engine}},\ }\href {https://doi.org/10.1016/j.ppnp.2022.104018}
  {\bibfield  {journal} {\bibinfo  {journal} {Prog. Part. Nucl. Phys.}\
  }\textbf {\bibinfo {volume} {129}},\ \bibinfo {pages} {104018} (\bibinfo
  {year} {2023})},\ \Eprint {https://arxiv.org/abs/2211.01050}
  {arXiv:2211.01050 [nucl-th]} \BibitemShut {NoStop}%
\bibitem [{\citenamefont {Thompson}\ \emph {et~al.}(2005)\citenamefont
  {Thompson}, \citenamefont {Quataert},\ and\ \citenamefont
  {Burrows}}]{Thompson:2004if}%
  \BibitemOpen
  \bibfield  {author} {\bibinfo {author} {\bibfnamefont {T.~A.}\ \bibnamefont
  {Thompson}}, \bibinfo {author} {\bibfnamefont {E.}~\bibnamefont {Quataert}},\
  and\ \bibinfo {author} {\bibfnamefont {A.}~\bibnamefont {Burrows}},\
  }\bibfield  {title} {\bibinfo {title} {{Viscosity and rotation in core -
  collapse supernovae}},\ }\href {https://doi.org/10.1086/427177} {\bibfield
  {journal} {\bibinfo  {journal} {Astrophys. J.}\ }\textbf {\bibinfo {volume}
  {620}},\ \bibinfo {pages} {861} (\bibinfo {year} {2005})},\ \Eprint
  {https://arxiv.org/abs/astro-ph/0403224} {arXiv:astro-ph/0403224}
  \BibitemShut {NoStop}%
\bibitem [{\citenamefont {Spruit}(2002)}]{Spruit:2001tz}%
  \BibitemOpen
  \bibfield  {author} {\bibinfo {author} {\bibfnamefont {H.~C.}\ \bibnamefont
  {Spruit}},\ }\bibfield  {title} {\bibinfo {title} {{Dynamo action by
  differential rotation in a stably stratified stellar interior}},\ }\href
  {https://doi.org/10.1051/0004-6361:20011465} {\bibfield  {journal} {\bibinfo
  {journal} {Astron. Astrophys.}\ }\textbf {\bibinfo {volume} {381}},\ \bibinfo
  {pages} {923} (\bibinfo {year} {2002})},\ \Eprint
  {https://arxiv.org/abs/astro-ph/0108207} {arXiv:astro-ph/0108207}
  \BibitemShut {NoStop}%
\bibitem [{\citenamefont {Kolomeitsev}\ and\ \citenamefont
  {Voskresensky}(2015)}]{Kolomeitsev15}%
  \BibitemOpen
  \bibfield  {author} {\bibinfo {author} {\bibfnamefont {E.~E.}\ \bibnamefont
  {Kolomeitsev}}\ and\ \bibinfo {author} {\bibfnamefont {D.~N.}\ \bibnamefont
  {Voskresensky}},\ }\bibfield  {title} {\bibinfo {title} {{Viscosity of
  neutron star matter and $r$-modes in rotating pulsars}},\ }\href
  {https://doi.org/10.1103/PhysRevC.91.025805} {\bibfield  {journal} {\bibinfo
  {journal} {Phys. Rev. C}\ }\textbf {\bibinfo {volume} {91}},\ \bibinfo
  {pages} {025805} (\bibinfo {year} {2015})},\ \Eprint
  {https://arxiv.org/abs/1412.0314} {arXiv:1412.0314 [nucl-th]} \BibitemShut
  {NoStop}%
\bibitem [{\citenamefont {Sagert}\ \emph {et~al.}(2009)\citenamefont {Sagert},
  \citenamefont {Fischer}, \citenamefont {Hempel}, \citenamefont {Pagliara},
  \citenamefont {Schaffner-Bielich}, \citenamefont {Mezzacappa}, \citenamefont
  {Thielemann},\ and\ \citenamefont {Liebend\"orfer}}]{Sagert2009}%
  \BibitemOpen
  \bibfield  {author} {\bibinfo {author} {\bibfnamefont {I.}~\bibnamefont
  {Sagert}}, \bibinfo {author} {\bibfnamefont {T.}~\bibnamefont {Fischer}},
  \bibinfo {author} {\bibfnamefont {M.}~\bibnamefont {Hempel}}, \bibinfo
  {author} {\bibfnamefont {G.}~\bibnamefont {Pagliara}}, \bibinfo {author}
  {\bibfnamefont {J.}~\bibnamefont {Schaffner-Bielich}}, \bibinfo {author}
  {\bibfnamefont {A.}~\bibnamefont {Mezzacappa}}, \bibinfo {author}
  {\bibfnamefont {F.-K.}\ \bibnamefont {Thielemann}},\ and\ \bibinfo {author}
  {\bibfnamefont {M.}~\bibnamefont {Liebend\"orfer}},\ }\bibfield  {title}
  {\bibinfo {title} {Signals of the qcd phase transition in core-collapse
  supernovae},\ }\href {https://doi.org/10.1103/PhysRevLett.102.081101}
  {\bibfield  {journal} {\bibinfo  {journal} {Phys. Rev. Lett.}\ }\textbf
  {\bibinfo {volume} {102}},\ \bibinfo {pages} {081101} (\bibinfo {year}
  {2009})}\BibitemShut {NoStop}%
\bibitem [{\citenamefont {Fischer}\ \emph {et~al.}(2018)\citenamefont
  {Fischer}, \citenamefont {Bastian}, \citenamefont {Wu}, \citenamefont
  {Baklanov}, \citenamefont {Sorokina}, \citenamefont {Blinnikov},
  \citenamefont {Typel}, \citenamefont {Klähn},\ and\ \citenamefont
  {Blaschke}}]{Fischer_2018}%
  \BibitemOpen
  \bibfield  {author} {\bibinfo {author} {\bibfnamefont {T.}~\bibnamefont
  {Fischer}}, \bibinfo {author} {\bibfnamefont {N.-U.~F.}\ \bibnamefont
  {Bastian}}, \bibinfo {author} {\bibfnamefont {M.-R.}\ \bibnamefont {Wu}},
  \bibinfo {author} {\bibfnamefont {P.}~\bibnamefont {Baklanov}}, \bibinfo
  {author} {\bibfnamefont {E.}~\bibnamefont {Sorokina}}, \bibinfo {author}
  {\bibfnamefont {S.}~\bibnamefont {Blinnikov}}, \bibinfo {author}
  {\bibfnamefont {S.}~\bibnamefont {Typel}}, \bibinfo {author} {\bibfnamefont
  {T.}~\bibnamefont {Klähn}},\ and\ \bibinfo {author} {\bibfnamefont {D.~B.}\
  \bibnamefont {Blaschke}},\ }\bibfield  {title} {\bibinfo {title} {Quark
  deconfinement as a supernova explosion engine for massive blue supergiant
  stars},\ }\href {https://doi.org/10.1038/s41550-018-0583-0} {\bibfield
  {journal} {\bibinfo  {journal} {Nature Astronomy}\ }\textbf {\bibinfo
  {volume} {2}},\ \bibinfo {pages} {980–986} (\bibinfo {year}
  {2018})}\BibitemShut {NoStop}%
\bibitem [{\citenamefont {Caprini}\ \emph {et~al.}(2020)\citenamefont
  {Caprini}, \citenamefont {Chala}, \citenamefont {Dorsch}, \citenamefont
  {Hindmarsh}, \citenamefont {Huber}, \citenamefont {Konstandin}, \citenamefont
  {Kozaczuk}, \citenamefont {Nardini}, \citenamefont {No}, \citenamefont
  {Rummukainen}, \citenamefont {Schwaller}, \citenamefont {Servant},
  \citenamefont {Tranberg},\ and\ \citenamefont {Weir}}]{Caprini2020}%
  \BibitemOpen
  \bibfield  {author} {\bibinfo {author} {\bibfnamefont {C.}~\bibnamefont
  {Caprini}}, \bibinfo {author} {\bibfnamefont {M.}~\bibnamefont {Chala}},
  \bibinfo {author} {\bibfnamefont {G.~C.}\ \bibnamefont {Dorsch}}, \bibinfo
  {author} {\bibfnamefont {M.}~\bibnamefont {Hindmarsh}}, \bibinfo {author}
  {\bibfnamefont {S.~J.}\ \bibnamefont {Huber}}, \bibinfo {author}
  {\bibfnamefont {T.}~\bibnamefont {Konstandin}}, \bibinfo {author}
  {\bibfnamefont {J.}~\bibnamefont {Kozaczuk}}, \bibinfo {author}
  {\bibfnamefont {G.}~\bibnamefont {Nardini}}, \bibinfo {author} {\bibfnamefont
  {J.~M.}\ \bibnamefont {No}}, \bibinfo {author} {\bibfnamefont
  {K.}~\bibnamefont {Rummukainen}}, \bibinfo {author} {\bibfnamefont
  {P.}~\bibnamefont {Schwaller}}, \bibinfo {author} {\bibfnamefont
  {G.}~\bibnamefont {Servant}}, \bibinfo {author} {\bibfnamefont
  {A.}~\bibnamefont {Tranberg}},\ and\ \bibinfo {author} {\bibfnamefont
  {D.~J.}\ \bibnamefont {Weir}},\ }\bibfield  {title} {\bibinfo {title}
  {Detecting gravitational waves from cosmological phase transitions with lisa:
  an update},\ }\href {https://doi.org/10.1088/1475-7516/2020/03/024}
  {\bibfield  {journal} {\bibinfo  {journal} {Journal of Cosmology and
  Astroparticle Physics}\ }\textbf {\bibinfo {volume} {2020}}\bibinfo  {number}
  { (03)},\ \bibinfo {pages} {024}}\BibitemShut {NoStop}%
\bibitem [{\citenamefont {Caprini}\ and\ \citenamefont
  {Figueroa}(2018)}]{Caprini2018}%
  \BibitemOpen
\bibfield  {number} {  }\bibfield  {author} {\bibinfo {author} {\bibfnamefont
  {C.}~\bibnamefont {Caprini}}\ and\ \bibinfo {author} {\bibfnamefont {D.~G.}\
  \bibnamefont {Figueroa}},\ }\bibfield  {title} {\bibinfo {title}
  {Cosmological backgrounds of gravitational waves},\ }\href
  {https://doi.org/10.1088/1361-6382/aac608} {\bibfield  {journal} {\bibinfo
  {journal} {Classical and Quantum Gravity}\ }\textbf {\bibinfo {volume}
  {35}},\ \bibinfo {pages} {163001} (\bibinfo {year} {2018})}\BibitemShut
  {NoStop}%
\bibitem [{\citenamefont {Auclair}\ \emph {et~al.}(2023)\citenamefont
  {Auclair}, \citenamefont {Bacon}, \citenamefont {Baker}, \citenamefont
  {Barreiro}, \citenamefont {Bartolo}, \citenamefont {Belgacem}, \citenamefont
  {Bellomo}, \citenamefont {Ben-Dayan}, \citenamefont {Bertacca}, \citenamefont
  {Besancon} \emph {et~al.}}]{Auclair2023}%
  \BibitemOpen
  \bibfield  {author} {\bibinfo {author} {\bibfnamefont {P.}~\bibnamefont
  {Auclair}}, \bibinfo {author} {\bibfnamefont {D.}~\bibnamefont {Bacon}},
  \bibinfo {author} {\bibfnamefont {T.}~\bibnamefont {Baker}}, \bibinfo
  {author} {\bibfnamefont {T.}~\bibnamefont {Barreiro}}, \bibinfo {author}
  {\bibfnamefont {N.}~\bibnamefont {Bartolo}}, \bibinfo {author} {\bibfnamefont
  {E.}~\bibnamefont {Belgacem}}, \bibinfo {author} {\bibfnamefont
  {N.}~\bibnamefont {Bellomo}}, \bibinfo {author} {\bibfnamefont
  {I.}~\bibnamefont {Ben-Dayan}}, \bibinfo {author} {\bibfnamefont
  {D.}~\bibnamefont {Bertacca}}, \bibinfo {author} {\bibfnamefont
  {M.}~\bibnamefont {Besancon}}, \emph {et~al.},\ }\bibfield  {title} {\bibinfo
  {title} {Cosmology with the laser interferometer space antenna},\ }\href@noop
  {} {\bibfield  {journal} {\bibinfo  {journal} {Living Rev. Relativ.}\
  }\textbf {\bibinfo {volume} {26}} (\bibinfo {year} {2023})}\BibitemShut
  {NoStop}%
\bibitem [{\citenamefont {Weir}(2018)}]{Weir2018}%
  \BibitemOpen
  \bibfield  {author} {\bibinfo {author} {\bibfnamefont {D.~J.}\ \bibnamefont
  {Weir}},\ }\bibfield  {title} {\bibinfo {title} {Gravitational waves from a
  first-order electroweak phase transition: a brief review},\ }\href
  {http://www.jstor.org/stable/44678718} {\bibfield  {journal} {\bibinfo
  {journal} {Philosophical Transactions: Mathematical, Physical and Engineering
  Sciences}\ }\textbf {\bibinfo {volume} {376}},\ \bibinfo {pages} {1}
  (\bibinfo {year} {2018})}\BibitemShut {NoStop}%
\bibitem [{\citenamefont {Ramsey-Musolf}(2020)}]{Ramsey-Musolf2020}%
  \BibitemOpen
  \bibfield  {author} {\bibinfo {author} {\bibfnamefont {M.~J.}\ \bibnamefont
  {Ramsey-Musolf}},\ }\bibfield  {title} {\bibinfo {title} {The electroweak
  phase transition: a collider target},\ }\href@noop {} {\bibfield  {journal}
  {\bibinfo  {journal} {J. High Energy Phys.}\ }\textbf {\bibinfo {volume}
  {2020}}\bibinfo  {number} { (9)}}\BibitemShut {NoStop}%
\bibitem [{\citenamefont {Mirón-Granese}(2021)}]{MironGranese2021}%
  \BibitemOpen
\bibfield  {number} {  }\bibfield  {author} {\bibinfo {author} {\bibfnamefont
  {N.}~\bibnamefont {Mirón-Granese}},\ }\bibfield  {title} {\bibinfo {title}
  {Relativistic viscous effects on the primordial gravitational waves
  spectrum},\ }\href {https://doi.org/10.1088/1475-7516/2021/06/008} {\bibfield
   {journal} {\bibinfo  {journal} {Journal of Cosmology and Astroparticle
  Physics}\ }\textbf {\bibinfo {volume} {2021}}\bibinfo  {number} { (06)},\
  \bibinfo {pages} {008}}\BibitemShut {NoStop}%
\bibitem [{\citenamefont {Sch{\"a}fer}\ and\ \citenamefont
  {Teaney}(2009)}]{Schafer2009}%
  \BibitemOpen
\bibfield  {number} {  }\bibfield  {author} {\bibinfo {author} {\bibfnamefont
  {T.}~\bibnamefont {Sch{\"a}fer}}\ and\ \bibinfo {author} {\bibfnamefont
  {D.}~\bibnamefont {Teaney}},\ }\bibfield  {title} {\bibinfo {title} {Nearly
  perfect fluidity: from cold atomic gases to hot quark gluon plasmas},\
  }\href@noop {} {\bibfield  {journal} {\bibinfo  {journal} {Rep. Prog. Phys.}\
  }\textbf {\bibinfo {volume} {72}},\ \bibinfo {pages} {126001} (\bibinfo
  {year} {2009})}\BibitemShut {NoStop}%
\bibitem [{\citenamefont {Trachenko}\ \emph {et~al.}(2021)\citenamefont
  {Trachenko}, \citenamefont {Brazhkin},\ and\ \citenamefont
  {Baggioli}}]{Trachenko2021}%
  \BibitemOpen
  \bibfield  {author} {\bibinfo {author} {\bibfnamefont {K.}~\bibnamefont
  {Trachenko}}, \bibinfo {author} {\bibfnamefont {V.}~\bibnamefont
  {Brazhkin}},\ and\ \bibinfo {author} {\bibfnamefont {M.}~\bibnamefont
  {Baggioli}},\ }\bibfield  {title} {\bibinfo {title} {Similarity between the
  kinematic viscosity of quark-gluon plasma and liquids at the viscosity
  minimum},\ }\href@noop {} {\bibfield  {journal} {\bibinfo  {journal} {SciPost
  Phys.}\ }\textbf {\bibinfo {volume} {10}} (\bibinfo {year}
  {2021})}\BibitemShut {NoStop}%
\bibitem [{\citenamefont {Misner}\ \emph {et~al.}(1973)\citenamefont {Misner},
  \citenamefont {{etc.}}, \citenamefont {Thorne},\ and\ \citenamefont
  {Wheeler}}]{Misner1973-mq}%
  \BibitemOpen
  \bibfield  {author} {\bibinfo {author} {\bibfnamefont {C.~W.}\ \bibnamefont
  {Misner}}, \bibinfo {author} {\bibnamefont {{etc.}}}, \bibinfo {author}
  {\bibfnamefont {K.~S.}\ \bibnamefont {Thorne}},\ and\ \bibinfo {author}
  {\bibfnamefont {J.~A.}\ \bibnamefont {Wheeler}},\ }\href@noop {} {\emph
  {\bibinfo {title} {Gravitation}}}\ (\bibinfo  {publisher} {W.H. Freeman},\
  \bibinfo {address} {New York, NY},\ \bibinfo {year} {1973})\BibitemShut
  {NoStop}%
\bibitem [{\citenamefont {Renzini}\ \emph {et~al.}(2022)\citenamefont
  {Renzini}, \citenamefont {Goncharov}, \citenamefont {Jenkins},\ and\
  \citenamefont {Meyers}}]{Renzini2022}%
  \BibitemOpen
  \bibfield  {author} {\bibinfo {author} {\bibfnamefont {A.~I.}\ \bibnamefont
  {Renzini}}, \bibinfo {author} {\bibfnamefont {B.}~\bibnamefont {Goncharov}},
  \bibinfo {author} {\bibfnamefont {A.~C.}\ \bibnamefont {Jenkins}},\ and\
  \bibinfo {author} {\bibfnamefont {P.~M.}\ \bibnamefont {Meyers}},\ }\bibfield
   {title} {\bibinfo {title} {Stochastic gravitational-wave backgrounds:
  Current detection efforts and future prospects},\ }\bibfield  {journal}
  {\bibinfo  {journal} {Galaxies}\ }\textbf {\bibinfo {volume} {10}},\ \href
  {https://doi.org/10.3390/galaxies10010034} {10.3390/galaxies10010034}
  (\bibinfo {year} {2022})\BibitemShut {NoStop}%
\bibitem [{\citenamefont {Melia}(2022)}]{Melia2022}%
  \BibitemOpen
  \bibfield  {author} {\bibinfo {author} {\bibfnamefont {F.}~\bibnamefont
  {Melia}},\ }\bibfield  {title} {\bibinfo {title} {The electroweak horizon
  problem},\ }\href
  {https://doi.org/https://doi.org/10.1016/j.dark.2022.101057} {\bibfield
  {journal} {\bibinfo  {journal} {Physics of the Dark Universe}\ }\textbf
  {\bibinfo {volume} {37}},\ \bibinfo {pages} {101057} (\bibinfo {year}
  {2022})}\BibitemShut {NoStop}%
\end{thebibliography}%
\end{document}